\documentclass[11pt]{article}
\usepackage[affil-it]{authblk} 
\usepackage[margin=1in]{geometry} 
\usepackage{subfigure,epsfig,floatflt,lscape,graphicx,pdflscape}
\usepackage[colorlinks=true,linkcolor=black,anchorcolor=black,citecolor=black,filecolor=black,menucolor=black,runcolor=black,urlcolor=black]{hyperref}
\usepackage{setspace,xr}
\usepackage{color,multirow,bm,enumitem,natbib}
\usepackage{xr}
\usepackage{amsmath}

\title{Identifying regions of inhomogeneities in spatial processes via an M-RA and mixture priors}
\author{Marco H. Benedetti$^{1}$ , 
Veronica J. Berrocal$^{2}$, and 
Naveen N. Narisetty$^{3}$}  
\affil{$^{1}$ Center for Injury Research and Policy, Nationwide Children's Hospital, Columbus, OH 43215
\\
$^{2}$ Department of Statistics, University of California, Irvine, CA 92697
\\
$^{3}$ Department of Statistics, University of Illinois at Urbana-Champaign, Urbana-Champaign, IL 61820}

\begin{document}
\maketitle

\label{firstpage}

\begin{abstract}
Soils have been heralded as a hidden resource that can be leveraged to mitigate and address some of the major global environmental challenges. Specifically, the organic carbon stored in soils, called Soil Organic Carbon (SOC), can, through proper soil management, help offset fuel emissions, increase food productivity, and improve water quality. As collecting data on SOC is costly and time consuming, not much data on SOC is available, although understanding the spatial variability in SOC is of fundamental importance for effective soil management. 

In this manuscript, we propose a modeling framework that can be used to gain a better understanding of the dependence structure of a spatial process by identifying regions within a spatial domain where the process displays the same spatial correlation range. To achieve this goal, we propose a generalization of the Multi-Resolution Approximation (M-RA) modeling framework of \citet{Katzfuss2017} originally introduced as a strategy to reduce the computational burden encountered when analyzing massive spatial datasets. 

 To allow for the possibility that the correlation of a spatial process might be characterized by a different range in different subregions of a spatial domain, we provide the M-RA basis functions weights with a two-component mixture prior with one of the mixture components a shrinking prior. We call our approach the \emph{mixture M-RA}.  Application of the mixture M-RA model to both stationary and non-stationary data shows that the mixture M-RA model can handle both types of data, can correctly establish the type of spatial dependence structure in the data (e.g. stationary vs not), and can identify regions of local stationarity. 
\end{abstract}

Keywords: Basis expansion; Gaussian process; inhomogeneities; local stationarity; mixture prior; M-RA; non-stationary covariance function.

\section{Introduction}\label{sec:intro}

Soils contain a massive proportion of the Earth system's carbon \citep{FAOreport}: within the first meters below surface, there is more organic carbon than in the atmosphere and terrestrial vegetation combined. This carbon, called soil organic carbon (SOC), is stored within soil organic matter, e.g. plant or animal residual matter, and is typically very dynamic due to its involvement in several biological, physical and chemical processes \citep{Bliss&2014}. For example, through root biomass decomposition, humification and particle detachment, SOC  gets deposited and distributed in soils and removed from the atmosphere, whereas, by acting as a membrane that filters pollutants, SOC plays a key role in water quality.
Human activity, along with erosion and SOC depletion, have contributed to a loss of SOC that varies not only with latitude but also locally. The adoption of recommended soil management practices, such as reduced tillage, manuring, etc., have the potential to transform soils into carbon pools, sequestering carbon from the atmosphere,  offsetting fossil fuel emissions, and improving agronomic yield and water quality. As such, SOC can provide a sustainable partial solution to three of the major global environmental challenges identified by the United Nations \citep{FAOreport}: climate change, desertification and food insecurity. 

Because of the biophysical and societal importance of SOC, since the Kyoto meeting, the Intergovernmental Panel on Climate Change has recommended that SOC stocks, i.e. the amount of SOC in a volume of soil, are carefully monitored  \citep{FAOreport}. However, since collection and measurement of SOC is costly and time consuming \citep{Goidts2007}, information on SOC is spatially sparse, with large disparities in terms of data availability across the planet. In addition, as  soil restorative and management measures have to be highly localized in order to be effective, there is a great interest in understanding the spatial variability and spatial dependence of SOC. This increased knowledge will result extremely useful when designing future sampling campaigns, which should be concentrated in regions with a fast-decaying spatial correlation and large prediction uncertainty.

As already indicated by \citet{Risser&2018}, SOC is characterized by a spatial dependence structure that is not homogeneous in space. 
The contribution of our paper is to present a novel statistical modeling approach that does not require a practitioner to determine a priori if a process is stationary or not, and can be used to model both stationary and non-stationary spatial processes. In the latter case, our model allows one to determine regions in which the spatial process is characterized by a similar range in the spatial correlation. This information is not readily available by other methods.
In fact, while tests to assess whether a process is stationary \citep{Bandyo2017,Jun2012}, isotropic \citep{Guan&2004} or symmetric \citep{Li2008,WellerHoeting2016} have been proposed in the literature, they only provide a global answer to the question of whether a spatial process is stationary or not, and they do not identify regions with similar strength in the spatial correlation. 

The decomposition of a spatial domain into areas with similar spatial dependence is one of the classical modeling approaches used to construct non-stationary covariance functions \citep{Sampson2010}, see for example \citet{Fuentes2001}, \citet{Kim&2005}, \citet{NottDunsmuir2002}, \citet{Pope&2018}, \citet{GramacyLee2008}, \citet{Konomi&2014}, and, more recently, \citet{Risser&2018} who leverage information in the covariates to define the domain segmentation. 
Our model adds to this literature on non-stationary spatial processes, and it breaks with previous approaches in that it provides an alternative, computationally less challenging way to determine regions of local stationarity.
Additionally, it does not  require specifying a priori the maximum number of possible segments, and it keeps computation rather feasible. 

We build on the Multi-Resolution Approximation (M-RA) model of \citet{Katzfuss2017} that provides an approximation to the original covariance function of a Gaussian spatial process via a linear combination of basis functions obtained by recursively implementing a predictive process approximation \citep{Banerjee&2008}. \citet{Katzfuss2017} noted that, in the M-RA modeling framework, the magnitude of the basis functions weights at each level is related to the strength of the spatial dependence in the data. 
Exploiting this intuition, we propose to modify the M-RA model to allow basis functions weights to be shrunk towards zero. Examining the behavior of the basis functions weights over space will provide us with information on whether the spatial dependence of the spatial process has the same strength across the domain. Because of the prior specification on the basis functions weights, we call our approach the \emph{mixture M-RA}. We show via simulation experiments that our modeling framework is flexible enough to accommodate both stationary and non-stationary data.

The remainder of the paper is organized as follows: in Section~\ref{sec:data} we introduce the motivating dataset and provide a description of the SOC data. In Section~\ref{sec:methods} we review the M-RA model and  present our modification of the M-RA approach to allow for the identification of regions of inhomogeneity in the range parameter. Section~\ref{sec:results} presents results relative to the application of our model to simulated data and SOC, while Section~\ref{sec:discussion} offers a discussion on limitations and future extensions of our model.  

\section{Soil Organic Carbon Data}\label{sec:data}
The Rapid Carbon Assessment (RaCA) project was a project initiated by the U.S. Department of Agriculture National Resource Conservation Service (NRCS) in 2010 with the goal of obtaining an up-to-date snapshot of the amount and spatial distribution of SOC across the conterminous United States (CONUS) under different land use and agricultural management. A secondary goal of the project was to provide useful information for subsequent modeling efforts that investigate changes in SOC as a result of climate change and conservation practices. 

Volumes of soils were extracted at tens of thousands of locations randomly selected within different land use/land cover classes, and the amount of SOC at depths of 5, 10, 20, 30, 50 and 100 centimeters (in Mg C ha$^{-1}$) were obtained using a Visible Near Infrared Spectrometer (VNIRS).  Here we focus on measurements of SOC at 100 cm of depth. The data are openly available in the \texttt{R} package \texttt{soilDB}.
A subset of these data, relative to the Great Lakes region, was previously analyzed by \citet{Risser&2018} who demonstrated that, contrarily to previous analyses of SOC that assumed second-order stationarity, SOC is indeed a non-stationary spatial process with varying correlation ranges across the Great Lakes region. 

\begin{figure}
    \centering
    \subfigure[]{\includegraphics[width=0.35\textwidth]{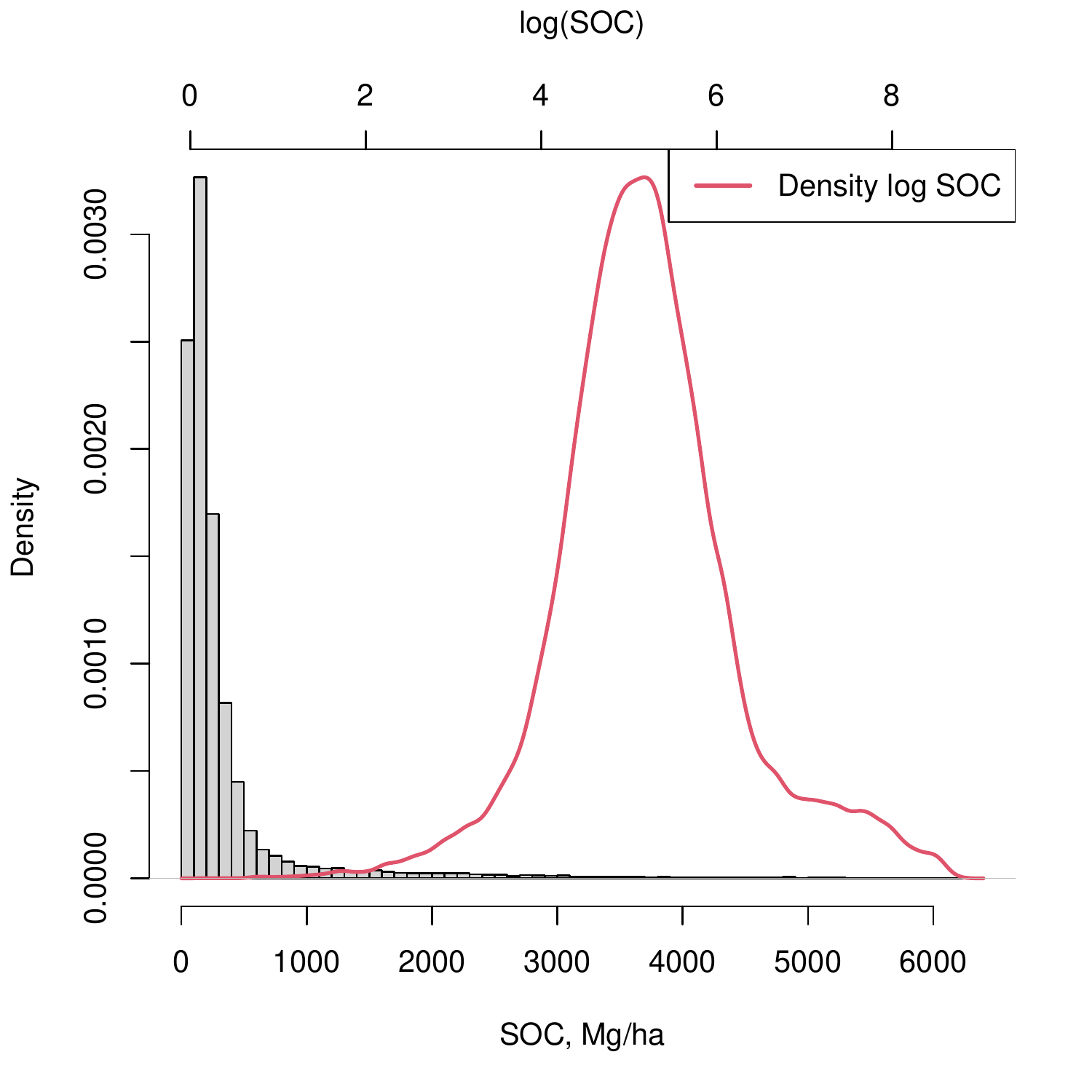}}  \qquad
    \subfigure[]{\includegraphics[width=0.45\textwidth]{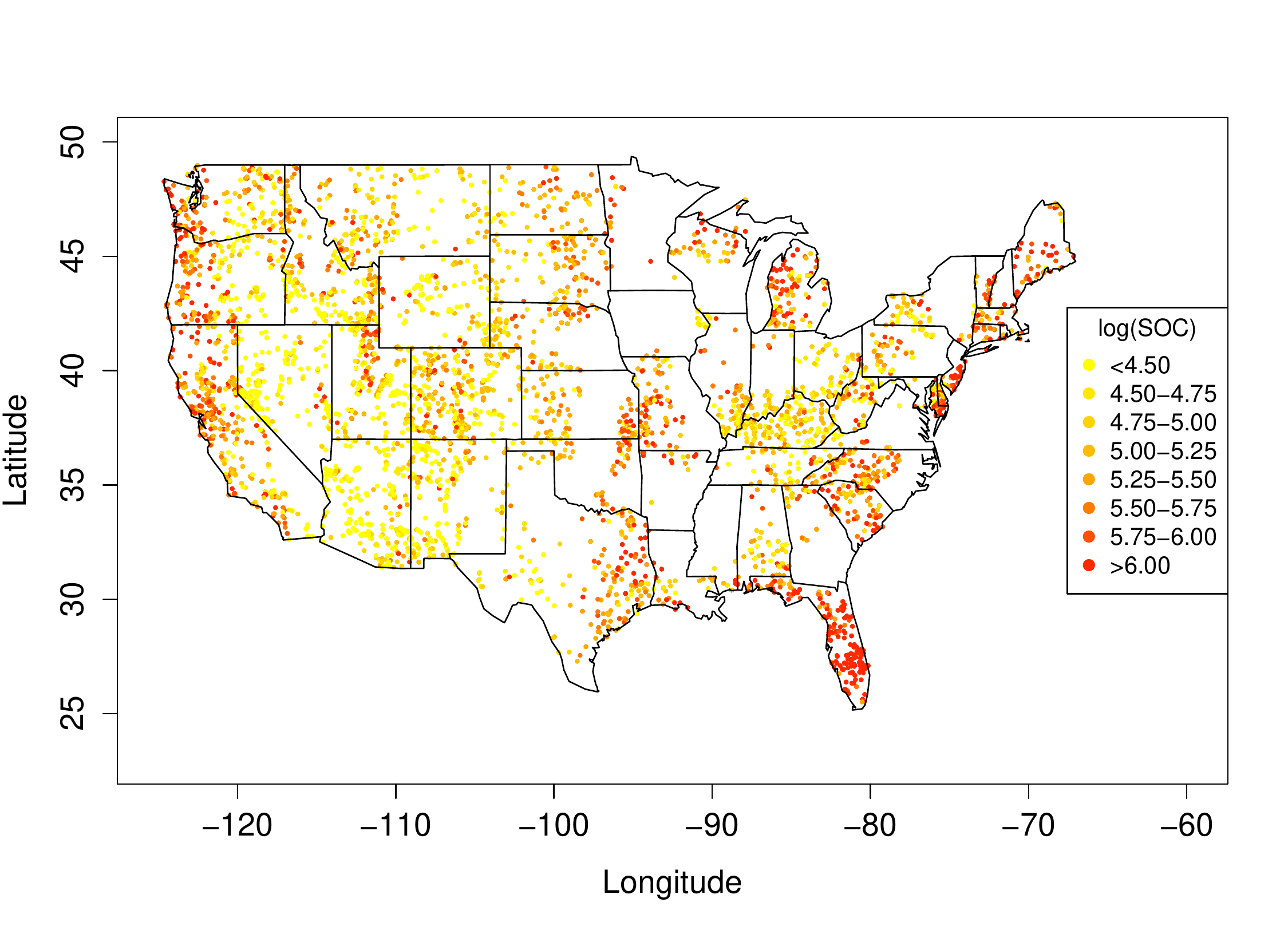}} \\
    \subfigure[]{\includegraphics[width=0.45\textwidth]{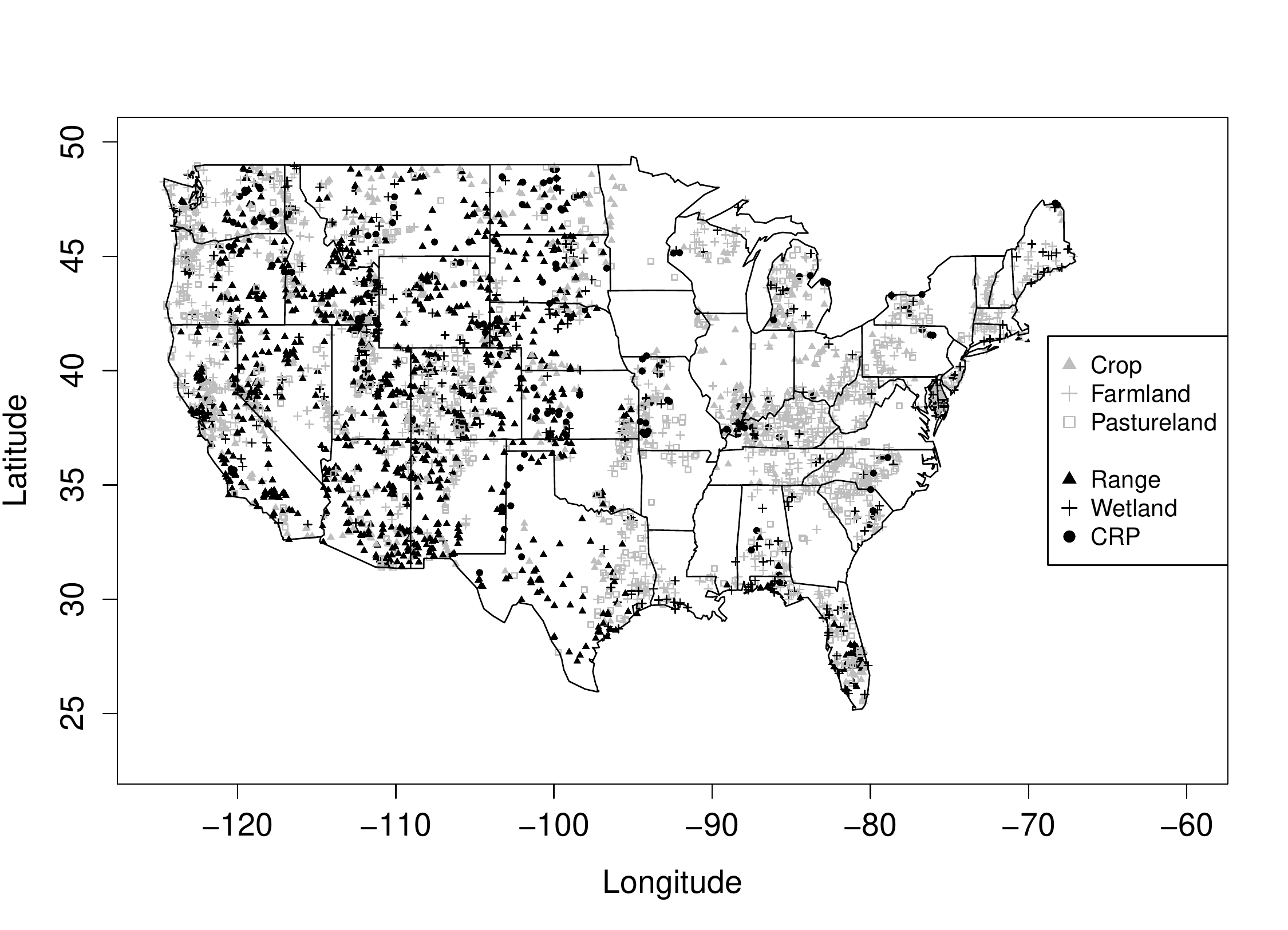}}  \qquad
    \subfigure[]{\includegraphics[width=0.45\textwidth]{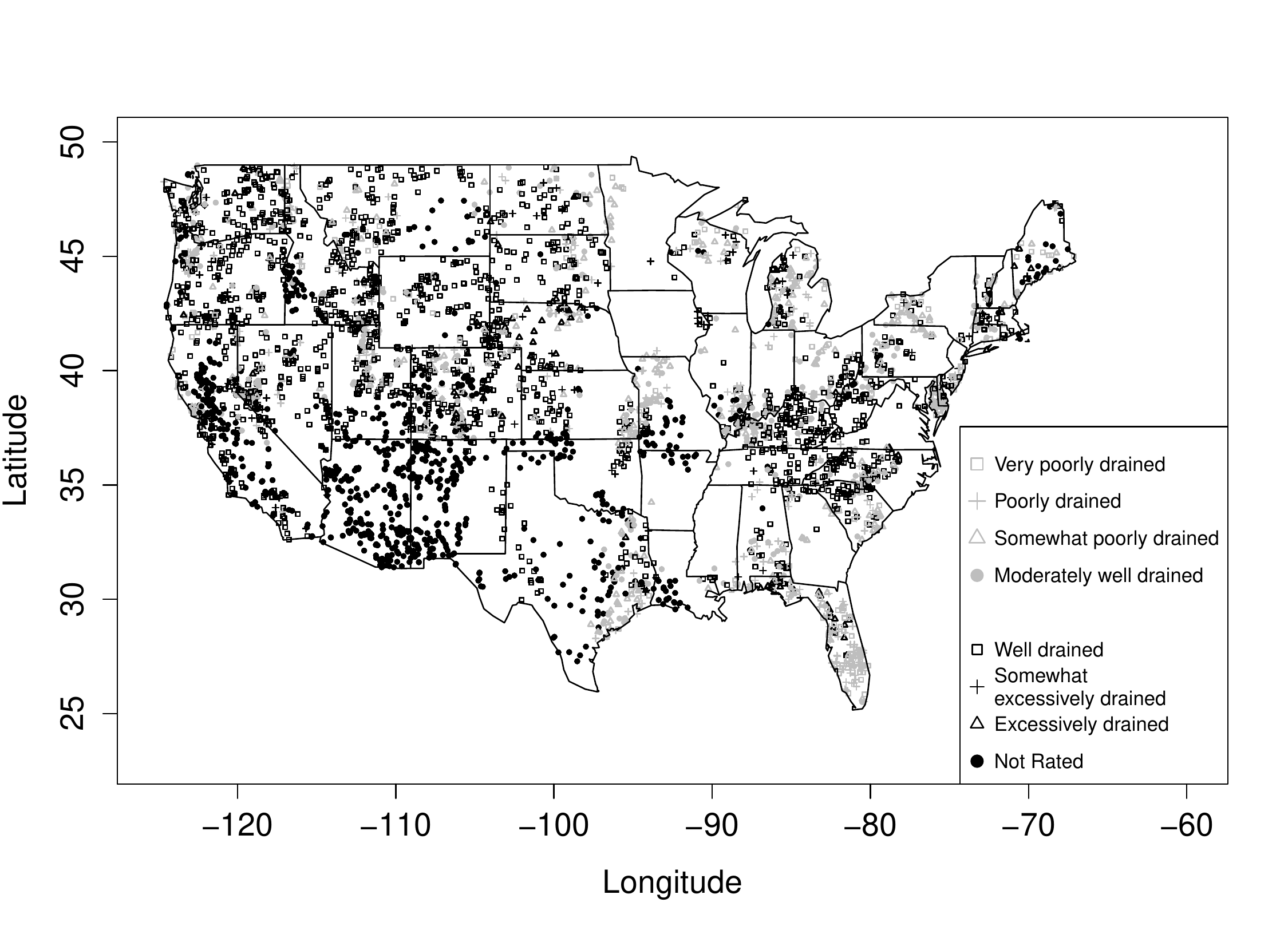}} \\
    \subfigure[]{\includegraphics[width=0.45\textwidth]{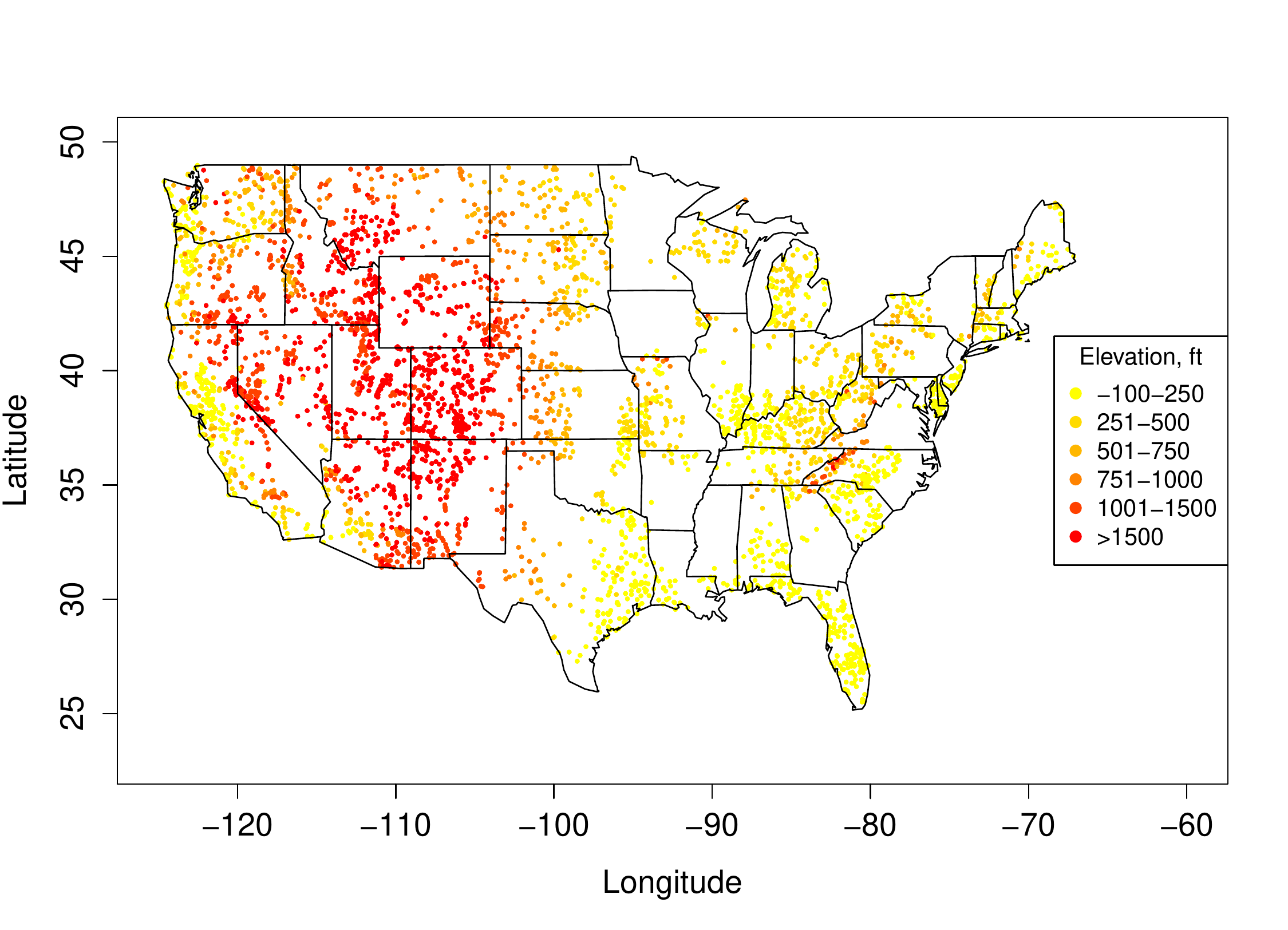}} \qquad
    \subfigure[]{\includegraphics[width=0.4\textwidth]{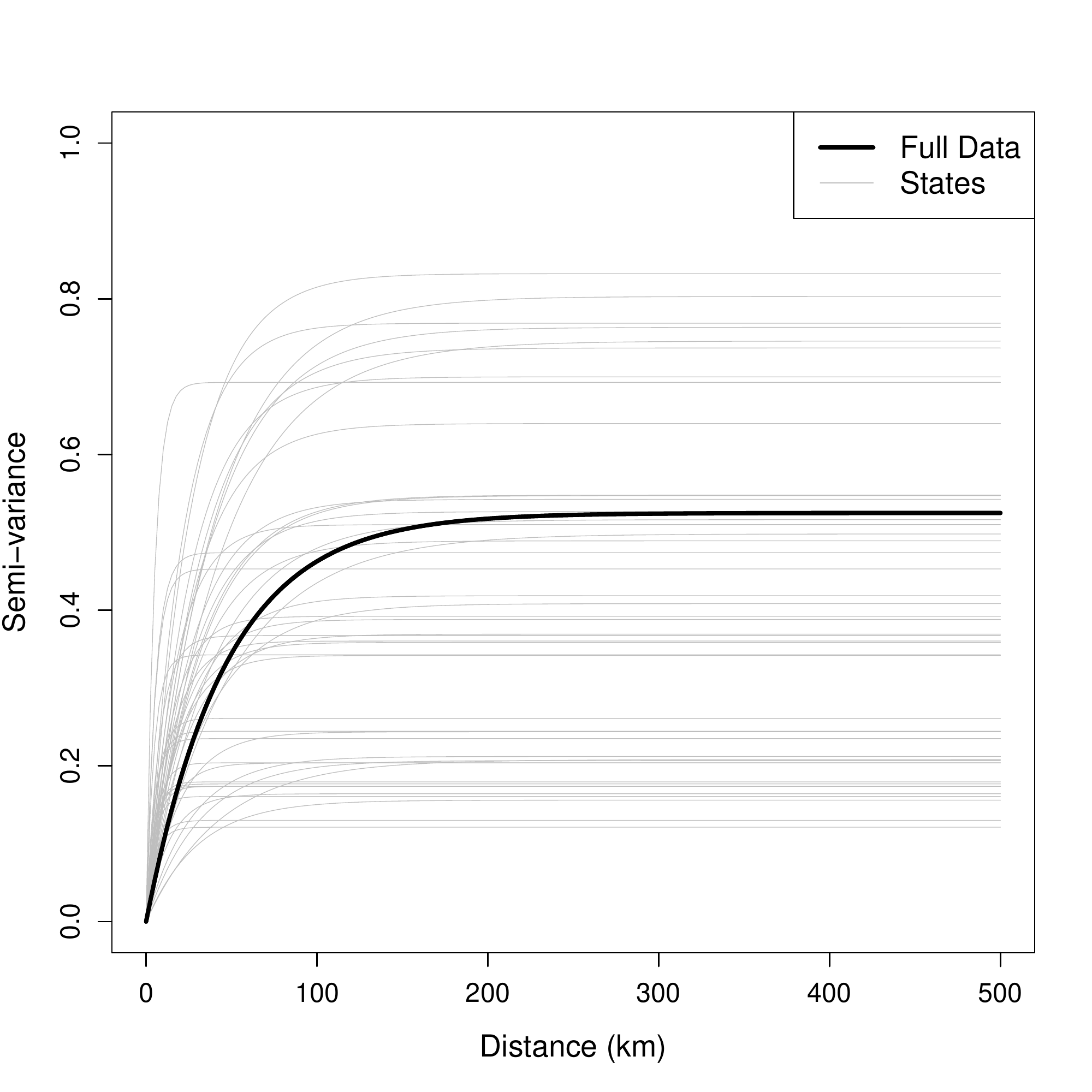}}
       \caption{Soil Organic Carbon (SOC) analysis. (a) Histogram of raw SOC values with overlaid a kernel density estimate of log SOC. (b)-(e) Spatial maps of: (b) log SOC; (c) land use/land cover; (d) drainage class; and (e) elevation. (f) Empirical semi-variogram of the residuals of log SOC on land use/land cover, drainage class and elevation when using data for each of the 48 CONUS states (thinner lines) and when using all the data (thicker line) across the CONUS.}
    \label{figure:exploratory}
\end{figure}

As Figure~\ref{figure:exploratory}(a) shows, raw SOC measurements are extremely right skewed, prompting us to work on the log scale, as did \citet{Risser&2018}. Log-transforming the data yields a distribution that is more symmetric, as Figure~\ref{figure:exploratory}(a) depicts. Across the CONUS, $\log$ SOC ranges from a minimum of 0.18 to a maximum of 8.71 log(Mg C ha$^{-1}$), with a mean of 5.07 and a standard deviation of 1.03 log(Mg C ha$^{-1}$).
In the remainder of the paper, with an abuse of terminology, we will use the expression ``SOC'' when referring to results that pertain to the analysis of $\log$ SOC. 

Building upon previous modeling efforts focused mostly on spatial prediction \citep{Risser&2018,Mishra2009}, in modeling the spatial process of SOC we expresse its mean function, $\mu(\mathbf{s})$, as a linear combination of three spatially-varying covariates - land use/land cover, drainage class, and elevation - known to influence levels of SOC. The first two, both categorical variables, are rather important from a geological point of view and refer, respectively, to: six different land use classifications according to the definition of the National Resources Inventory (namely: wetland, crop, pastureland, rangeland, forestland and land under a Conservation Reserve Program or CRP), and eight different classes that refer to the ``frequency and duration of wet periods under conditions similar to those under which the soil developed'' (NRCS Soil Survey Manual). Over 40\% of the SOC samples were taken in locations that are used primarily as pasture or for grazing by animals (26.0\% range and 16.8\% pastureland), while 26.4\% and 17.0\% of the soil samples refer to farmland and crop, respectively. SOC sites are also characterized by good drainage: about 50\% of the SOC samples were collected in well drained land (9.2\% moderately well drained, 40.9\% well drained and 
2.3\% excessively drained), whereas about 20\% of the sites were either poorly drained (8.2\% somewhat poorly drained, 8.7\% poorly drained and 4.7\% very poorly drained) or not rated (22.7\%).

Spatial plots of SOC, land use/land cover, drainage class, and elevation at the 20,087 locations in the CONUS, where measurements of all variables are available, are displayed in Figure~\ref{figure:exploratory}(b) through (e). Inspection of these maps indicate that SOC tends to be lower at higher altitudes, and higher in poorly drained locations. However, none of these covariates fully explain the spatial variability in SOC: as Table~\ref{tbl:linmodSOC} reports, a linear regression model of SOC on the three covariates yields a residual standard error of 0.88 log(Mg C ha$^{-1}$). 

To examine whether SOC is a stationary spatial process, we generate empirical semi-variograms of the residuals of the above-mentioned linear regression model for each of the 48 conterminous US states. The variograms, presented in Figure~\ref{figure:exploratory}(f), highlight the inhomogeneity in the strength of the spatial dependence in SOC across the CONUS, as also noted by \citet{Risser&2018}.
As we recognize that the 48 states determine a partition of the CONUS based on administrative boundaries, our goal in the rest of this paper is to develop a statistical modeling framework that allows us to identify regions with a similar spatial dependence structure in SOC.

\section{Methods}\label{sec:methods}
\subsection{The Multi-Resolution Approximation (M-RA)}\label{sec:MRA}
In this section, we provide a brief review of the M-RA approach. Interested readers are referred to \citet{Katzfuss2017} for additional details. In the following, we adopt a notation that is slightly different from that used by the former, particularly with respect to the subscripts used to index domain partitions and levels. 

Let $y(\mathbf{s}), \mathbf{s} \in \mathcal{S}$ denote a spatial process in $\mathcal{S}$ observed at locations $\mathbf{s}_1, \mathbf{s}_2, \ldots, \mathbf{s}_n$.  
Using a geostatistical modeling approach \citep{Banerjee&2004}, we express $y(\mathbf{s})$ as \vspace{-1ex}
\begin{equation}
y(\mathbf{s}) = \mu(\mathbf{s}) + w(\mathbf{s}) + \epsilon(\mathbf{s}) \qquad \epsilon(\mathbf{s}) \stackrel{iid}{\sim} N(0,\tau^2),
\label{eq:ch2geo}
\end{equation}
where $\mu(\mathbf{s})$ denotes the mean, or large scale spatial trend in $y(\mathbf{s})$, $w(\mathbf{s})$ indicates spatial random effects, and $\epsilon(\mathbf{s})$ denotes an independent error process, independent of $w(\mathbf{s})$. We will often refer to (\ref{eq:ch2geo}) as a Kriging model, using the expression Bayesian Kriging model if (\ref{eq:ch2geo}) is fit within a Bayesian framework.

In (\ref{eq:ch2geo}), the spatial process $w(\mathbf{s})$ is taken to be a mean-zero Gaussian process with covariance function $C_{w}(\mathbf{s}, \mathbf{s}^\prime;\boldsymbol{\theta})$, where $\boldsymbol{\theta}$ represents a vector of covariance parameters that, in the case of a stationary, isotropic covariance function, includes the marginal variance ($\sigma^2$) and the range parameter ($\phi$). Our $\boldsymbol{\theta}$ does not include a nugget effect ($\tau^2$) since that part of variability in the data is already accounted for by the term $\epsilon(\mathbf{s})$.
When $C_{w}(\mathbf{s}, \mathbf{s}^\prime;\boldsymbol{\theta})$ is the Mat\'{e}rn covariance function,   
$\boldsymbol{\theta}$ also includes  a smoothness parameter $\nu$. 

Using the M-RA framework, the spatial process $w(\mathbf{s})$ can be approximated by $\tilde{w}_{M}(\mathbf{s})$, defined as $\tilde{w}_{M}(\mathbf{s}) = \mathbf{B}(\mathbf{s}) \boldsymbol{\eta}$ with $\mathbf{B}(\mathbf{s})$ matrix of basis functions up to level $M$ evaluated at $\mathbf{s}$ and $\boldsymbol{\eta}$ set of basis functions weights. Replacing $w(\mathbf{s})$ with $\tilde{w}_{M}(\mathbf{s})$ into (\ref{eq:ch2geo}) leads to the \textit{M-RA model}
\begin{equation} 
y(\mathbf{s}) =  \mu(\mathbf{s}) + \tilde{w}_{M}(\mathbf{s}) + \epsilon(\mathbf{s})  =  \mu + \mathbf{B}(\mathbf{s}) \boldsymbol{\eta} + \epsilon(\mathbf{s}), \qquad \epsilon(\mathbf{s}) \stackrel{iid}{\sim} N(0,\tau^2),
\label{mramodel}
\end{equation}
which provides great computational efficiency for large dimensional spatial data as illustrated in \citet{Katzfuss2017}.

The basis functions $\mathbf{B}(\mathbf{s})$ are defined by recursively partitioning the spatial domain, introducing a new set of knots within each new partition and using a predictive process approximation each time.  
More precisely, at level $0$, $r$ knots are placed on the entire domain. No particular placement of the $r$ knots is suggested, although placing them on an equidistant grid is probably the most convenient and easy-to-implement choice. We indicate with $\mathcal{Q}^{(0)}$ the set of $r$ knots introduced at level $0$. Using this first set of knots $\mathcal{Q}^{(0)}$, the original process $w(\mathbf{s})$ is approximated using the predictive process $\tau_{0}(\mathbf{s}):= E\left[ w(\mathbf{s}) | w(\mathbf{Q}^{(0)}) \right]$ where $w(\mathcal{Q}^{(0)})$ denotes the $r$-dimensional realization of the spatial process $w(\mathbf{s})$ at the knots' locations. After this initial approximation at level $0$, at level $1$ the spatial domain is subdivided into $J$ non-overlapping subregions, and $r$ new knots are placed in each new subregion (see Web Figure 1 in Web Appendix A for an illustration). We indicate with $\mathcal{Q}^{(1)}$ the set of $J\cdot \! r$ knots introduced at level $1$. As for level 0, the knots in $\mathcal{Q}^{(1)}$ are in turn used to construct the predictive process approximation, $\tau_{1}(\mathbf{s})$, to the remainder process, $\delta_{1}(\mathbf{s})$, obtained at level $0$ and defined as $\delta_{1}(\mathbf{s}):= [w(\mathbf{s})-\tau_{0}(\mathbf{s})]$, where $[\cdot]$ superimposes independence across subregions. We note that $J$ and $r$ do not need to be equal at each level, but it is assumed for convenience of notation. In addition, the knots do not need to lay on a grid at each level: \citet{Katzfuss2017} uses the observation locations as knots in the final level of the M-RA. This procedure of partitioning, introducing knots, and approximating the remainder term $\delta_{m}(\mathbf{s})$ with its predictive process approximation $\tau_{m}(\mathbf{s})$ is repeated $M$ times leading to the following $M$-level M-RA approximation $w_{M}(\cdot)$ to $w(\mathbf{s})$: 
\begin{equation}
w_{M}(\mathbf{s}) = \tau_{0}(\mathbf{s}) + \tau_{1}(\mathbf{s}) + \ldots + \tau_{M}(\mathbf{s}) + \delta_{M+1}(\mathbf{s}) \equiv \tilde{w}_{M}(\mathbf{s}) + \delta_{M+1}(\mathbf{s}), \qquad \mathbf{s}\in \mathcal{S},
\label{eq:Mlevelapprox}
\end{equation}
with $\delta_{M+1}(\mathbf{s})$ remainder at level $M$.

By construction, the individual terms in (\ref{eq:Mlevelapprox}) are mutually independent processes as also are the remainder processes, $\delta_{1}(\mathbf{s})$, $\delta_{2}(\mathbf{s})$, $\ldots$, $\delta_{M+1}(\mathbf{s})$, with the latter independent across subregions at the corresponding level, e.g. $\delta_{1}(\mathbf{s})$ is independent across the $J$ subregions introduced at level $1$. This leads to a convenient block-diagonal covariance matrix structure for the basis functions weights. For a sufficiently large $M$, it is safe to assume that the remainder term $\delta_{M+1}(\mathbf{s})$ is not spatially correlated as the spatial dependence structure in $y(\mathbf{s})$ has been captured by $\tilde{w}_{M}(\mathbf{s})$. 

As at each level $m$, the predictive process $\tau_{m}(\mathbf{s})$ can be rewritten as a basis function expansion (see \citet{Banerjee&2008}), then it follows 
\begin{equation}
\tilde{w}_{M}(\mathbf{s})= \sum_{m=0}^{M} \sum_{j=1}^{J^m} \boldsymbol{b}_{m,j}(\mathbf{s}) \boldsymbol{\eta}_{m,j},
\label{wMbasis}
\end{equation}
where the sum is taken over partitions and levels (see \citet{Katzfuss2017} for more details). In (\ref{wMbasis}), $\mathbf{b}_{m,j}(\mathbf{s})$ denotes the set of basis functions corresponding to the $j$-th partition and the $m$-th level evaluated at $\mathbf{s}$, while $\boldsymbol{\eta}_{m,j}$ is the $r$-dimensional vector of basis functions weights in the $j$-th partition of the $m$-th level. 

The form of the basis functions $\mathbf{b}_{m,j}(\mathbf{s})$ is completely determined by the choice of the  covariance function $C_{w}(\cdot, \cdot)$ for the spatial random effects $w(\mathbf{s})$ in (1). For a given covariance function, the basis functions can be defined as follows (see \citet{Banerjee&2008} and \citet{Katzfuss2017} for more details): let $\mathcal{Q}^{(m,j)}$ denote the set of $r$ knots in the $j$-th partition at the $m$-th level, $m = 0,...,M, j = 1,...,J^m$, with $\mathcal{Q}^{(m)} = \bigcup_{j=1}^{J^{m}} \mathcal{Q}^{(m,j)}$, then the basis functions and the prior covariance of the basis functions weights are defined by the following recursive formulas: 
\begin{eqnarray}
v_0(\mathbf{s}_1,\mathbf{s}_2) & = & C_{w}(\mathbf{s}_1,\mathbf{s}_2;\boldsymbol{\theta}) \nonumber \\
\mathbf{b}_{m,j}(\mathbf{s}) & = & v_m( \mathbf{s}, \mathcal{Q}^{(m,j)}) \nonumber  \\
\mathbf{K}^{-1}_{m,j} & = & v_m(\mathcal{Q}^{(m,j)}, \mathcal{Q}^{(m,j)})  \label{eq:mraweights} \\
v_{m+1}(\mathbf{s}_1,\mathbf{s}_2) & = &  \begin{cases}
0, \text{ if } \mathbf{s}_1 \text{ and } \mathbf{s}_2 \text{ are in different regions at resolution $m$} \nonumber \\
  v_m(\mathbf{s}_1,\mathbf{s}_2) 
-\mathbf{b}_{m,j}(\mathbf{s}_1)'\mathbf{K}_{m,j}\mathbf{b}_{m,j}(\mathbf{s}_2), \text{ otherwise}    \nonumber
\end{cases} 
\end{eqnarray}
for any $\mathbf{s}_1, \mathbf{s}_2 \in \mathcal{S}$. For every $m$ and $j$, $\mathbf{K}_{m,j}$ is a $r\times r$ covariance matrix, and 
 \begin{equation}
 \boldsymbol{\eta}_{m,j} \sim N_{r}(\mathbf{0}, \mathbf{K}_{m,j}).
\label{eq:mradist}
\end{equation}
Replacing (\ref{wMbasis}) into (\ref{mramodel}) yields
\begin{equation}
y(\mathbf{s}) = \mu + \sum_{m=0}^{M} \sum_{j=1}^{J^m} \boldsymbol{b}_{m,j}(\mathbf{s}) \boldsymbol{\eta}_{m,j} + \epsilon(\mathbf{s}) ,  \qquad \epsilon(\mathbf{s})\stackrel{iid}{\sim} N(0,\tau^2).
\label{eq:modellev1}
\end{equation}

\subsection{The Mixture M-RA}\label{sec:ourmodel} 
To allow for the possibility that a spatial process is characterized by a spatial correlation with a different range parameter in different subregions, we propose to slightly change the prior distribution on the basis functions weights. Specifically, rather than specifying a multivariate normal prior, we provide them with a prior that allows them to be shrunk to zero from a level $\tilde{M} < M$ onward in certain subregions, if needed. Regions where the basis function weights are shrunk to zero correspond to segments of the spatial domain where the spatial process is characterized by a slowly-decaying spatial correlation.

Various alternatives are possible to shrink the basis functions weights to 0: a spike and slab prior \citep{MitchellBeauchamp1988,IshwaranRao2005}, nonlocal priors \citep{JohnsonRossell2012}, stochastic search variable selection \citep{GeorgeMcCulloch1993,GeorgeMcCulloch1997}, or empirical Bayes variable selection \citep{GeorgeFoster2000}. Here, given the large number of parameters, for computational convenience, we elect to use the method proposed by \citet{NarisettyHe2014} for Bayesian variable selection. This will enable us to retain the hierarchical structure of the basis functions weights, shrinking to 0 all the basis functions weights nested within a given subregion and level, while maintaing desirable properties for shrinkage. 
\newline Using the prior distribution in (\ref{eq:mradist}) as a starting point, we specify the following mixture prior: for $m=0,\ldots,M$ and $j=1,\ldots,J^{m}$  
\begin{equation}
\boldsymbol{\eta}_{m,j} \sim p_{m}N_r(0,\mathbf{K}_{m,j}) + (1-p_{m}) N_r(0,\mathbf{K}_{m,j}/L),
\label{eq:priorweights}
\end{equation}
where $L$ is a fixed large constant. The parameter $0 \le p_m \le 1$ in (\ref{eq:priorweights}) indicates the probability that $\boldsymbol{\eta}_{m,j}$ is not shrunk to $0$ and thus it is \emph{active}. We call this model the \emph{mixture M-RA}. In fitting the mixture M-RA model to data, we tune $L$ within the burn-in period of the Markov Chain Monte Carlo (MCMC) algorithm, keeping it fixed at the selected value for the remainder of the MCMC iterations. In future implementations, $L$ could be seen as an additional parameter in the model for which a prior distribution may be considered.


A characteristic distinction of our prior specification on the $\boldsymbol{\eta}_{m,j}$'s compared to the commonly used spike and slab Gaussian priors is that the covariance matrix of both the spike and slab distributions is proportional to $\mathbf{K}_{m,j}$ instead of being equal to the identity matrix. This is to reflect the dependence structure in the $\boldsymbol{\eta}_{m,j}$'s. 

The grouping-preservation character of our model specification can be seen in the following Bayesian hierarchical formulation. Let $Z_{m,j}$, for $m=0,\ldots,M$, $j =1,\ldots,J^{m}$ denote binary latent variables, then our model could be re-expressed as: 
\begin{eqnarray}
&y(\mathbf{s}) =  \mu(\mathbf{s}) + \sum_{m=0}^{M} \sum_{j=1}^{J^{m}} \mathbf{b}_{m,j} \boldsymbol{\eta}_{m,j}+  \epsilon(\mathbf{s})   \qquad   \epsilon(\mathbf{s}) \stackrel{iid}{\sim} N(0, \tau^2) \nonumber \vspace{1ex} \\
& \hspace{5ex} \boldsymbol{\eta}_{m,j} \mid Z_{m,j}=1  \sim  N_r(0,\mathbf{K}_{m,j})   \qquad  \boldsymbol{\eta}_{m,j}\mid Z_{m,j}=0  \sim   N_r(0,\mathbf{K}_{m,j}/L) \nonumber \\
&Z_{m,j} \mid (Z_{m-1,j^*}=1, ~p_m)  \sim \text{Bernoulli} (p_m); \quad p_m =  \rho^m; \quad \rho   \sim    \mbox{Beta} (\alpha_{\rho},\beta_{\rho}) \label{eq:priorweights2} \\
&   \qquad P(Z_{m,j}=1 \mid Z_{m-1,j^*}=0)  =  0   \label{eq:priorweights3}  
\end{eqnarray}
where $j^*$ is the partition in level $m-1$ that contains the $j$-th partition at the $m$-th level.

In (\ref{eq:priorweights2}), $p_m$ represents the probability that a set of basis functions weights in the $m$-th level belongs to the first component of the mixture prior and is not shrunk. Since we expect that for $m=0$, there is residual spatial correlation to be captured, all sets of weights will belong to the first mixture component, whereas at higher levels, with almost no residual spatial correlation left, they are more likely to be drawn from the second component of the mixture prior. In light of this, for model parsimony, we set $p_m$ to be equal to $\rho^m$, with $\rho\in (0,1)$. A more general framework will define $p_m=\rho^{cm}$ with a positive parameter $c$ to be estimated, or it will keep all the $p_m$'s distinct while still monotone decreasing in $m$. To set the weights at a given level to be zero if the weights in the previous level are zero, we add \eqref{eq:priorweights3} to our model specification. We note that our conditional prior specification in (\ref{eq:priorweights2}) and (\ref{eq:priorweights3}) is a special case of the priors proposed by \citet{Taylor-Rodriguez&2016} for variable selection in the presence of heredity constraints, and we observe that (\ref{eq:priorweights2}) and (\ref{eq:priorweights3}) impose a strong heredity constraint.

We complete the specification of our model by providing priors to all the remaining model parameters. 
Choosing a stationary Mat\'{e}rn covariance function for $C_{w}(\mathbf{s}, \mathbf{s}^\prime; \boldsymbol{\theta})$, the covariance parameter $\boldsymbol{\theta}$ is given by $(\sigma^2, \phi, \nu)^\prime$. 
We place Inverse Gamma priors on the residual variance, or nugget effect, $\tau^2$, and on the marginal variance $\sigma^2$ of $w(\mathbf{s})$. We choose hyperparameters $\alpha_{\tau^2}$, $\beta_{\tau^2}$ and $\alpha_{\sigma^2}$, $\beta_{\sigma^2}$, corresponding to the shape and rate parameter, respectively, so that the priors on $\tau^2$ and $\sigma^2$ are both weakly informative. Conversely, we specify a weakly informative Gamma$(0.001,0.001)$ prior  on the range parameter $\phi$, while we place a Uniform prior on the interval $(0,2)$ on the smoothness parameter $\nu$. Finally, assuming $\mu(\mathbf{s})\equiv \mu$, we place a vague mean-zero Normal prior on $\mu$.  In situations where $\mu(\mathbf{s})$ is modeled as a linear function of (spatial) covariates, the regression coefficients $\boldsymbol{\beta}$ are provided with flat prior distributions. We clarify that, in spite of using some improper priors on the  parameters for the mean function, because our model specifies a Gaussian likelihood, our joint posterior distribution is proper. 
%
\subsection{Posterior Inference} \label{sec:postinf}
 
We fit our model within a Bayesian framework, approximating the posterior distribution using an MCMC algorithm.
The algorithm includes Gibbs sampling steps to generate posterior samples for the constant mean $\mu$, the basis functions weights $\boldsymbol{\eta}_{m,j}$, the nugget effect $\tau^2$, and the auxiliary binary variables $Z_{m,j}$.
Metropolis-Hastings steps are used to generate posterior samples of the parameter $\rho$ that defines the probabilities $p_{m}$, and of the covariance parameters $\sigma^2$, $\phi$ and $\nu$. Specifically, to sample $\rho$ we use a uniform proposal distribution bounded between 0 and 1 and centered at its current value in the MCMC algorithm. Similarly, we use uniform proposals to sample $\sigma^2$, $\phi$ and $\nu$. In all cases, the widths of the uniform proposals are adaptively adjusted every 100 iterations until burn-in to achieve an acceptance rate of approximately 25\%. Details on the adaptive MCMC algorithm are provided in Web Appendix C.

Although in the current implementation of the mixture M-RA we do not place a prior on $L$, we determine its value within the MCMC algorithm. Specifically, we start the algorithm with a large value for $L$ for which we expect few to no basis functions weights to be drawn from the shrinkage prior. We typically use 1,000 as initial value for $L$ based on the results obtained in Simulation study 1, which we discuss in Section~\ref{sec:results}. We monitor the behavior of the basis functions weights during the burn-in period by computing the average of the sampled binary auxiliary variables $Z_{m,j}$ every 1,000 iterations, for $m=0,\ldots,M$ and $j=1,\ldots,J^{m}$. Expecting shrinkage of the basis functions weights mostly at the higher levels, we focus on monitoring the average of the $Z_{M,j}$'s, reducing the value of $L$ by 50\% if the average of the sampled $Z_{M,j}$'s is greater than 0.95, while keeping it fixed otherwise. We continue monitoring the basis functions weights, decreasing the value of $L$, if warranted, during burn-in until we do not see further significant changes in the proportion of basis functions weights being shrunk to zero. We then keep $L$ fixed for the rest of the MCMC iterations. More details on the MCMC algorithm along with the pseudocode are available in Web Appendix C.
 
\section{Results}\label{sec:results}

We now present results of the application of the mixture M-RA model to simulated data and to observations of Soil Organic Carbon in the CONUS. \vspace{-1ex}

\subsection{Simulation results}\label{sec:simres}
To gain a better understanding of the mixture M-RA model, we designed simulation studies with the goals of:
\begin{enumerate}
\item understanding the role of $L$ in the estimation of the basis functions weights, and the magnitude of $L$ needed to allow shrinkage to zero of the basis functions weights, when a process is truly locally stationary;
\item evaluating whether the mixture M-RA model can identify regions of local stationarity if they indeed exist, even in case of model mis-specification; and
\item determining whether shrinkage of the basis functions weights is needed to identify regions of non-stationarity.
\end{enumerate}

To achieve these objectives we considered the following simulations:
\begin{itemize}
\item \textit{Simulation study 1}: data were generated according to the M-RA model in (\ref{eq:modellev1}) with some basis functions weights $\boldsymbol{\eta}_{m,j}$ set equal to $0$; and
\item \textit{Simulation study 2}: data were generated on the unit square and non-stationarity was obtained by introducing two mean-zero stationary spatial processes with different spatial correlation ranges, each operating on one half of the square. 
\end{itemize}

For each simulation study, we generated multiple replicates: 50 for simulation study 1 and 30 for simulation study 2. Posterior inference is based on samples yielded from an MCMC algorithm whose convergence was assessed using various diagnostics, including trace plots, and Geweke's \citep{Geweke1992} and Raftery and Lewis' diagnostics \citep{Raftery1992}.  Unless otherwise noted, results are averaged across the multiple realizations. Here we report and discuss results for simulation studies 1 and 2; results for additional simulation studies investigating other properties of the mixture M-RA are available in Web Appendix D.

\subsubsection{Simulation Study 1}\label{sec:sim1res}
We generated data at $n = 756$ random locations in $\mathcal{S}$=[0,1] $\times$ [0,1] fifty times according to (\ref{eq:modellev1}) where $\mu=0$ and $\tau^2=0.05$. Basis functions weights were drawn either from the distribution in (\ref{eq:mradist}) with covariance matrix implied by a stationary Mat\'{e}rn covariance function with $\boldsymbol{\theta} = (\sigma^2, \phi,\nu)^\prime = (1,0.1,1.0)$, $M=3$, $J = 4$ and $r=9$, or were set equal to 0. Zero-valued function weights were located in the upper right and lower left quadrants (see Web Figure 6 in Web Appendix D).

To each of the 50 datasets, we fitted our mixture M-RA model using $M=3$, $J = 4$ and $r=16$. As the goal of this simulation study is to determine whether our model is capable to identify regions with different strengths of spatial dependence, in fitting the mixture M-RA model we did not estimate $\boldsymbol{\theta}$, rather we kept it fixed at its true value. However, we varied the value of $L$ and we used six different ones: $L$ =10, 25, 50, 100, 200, and 10,000. 

Table~\ref{tbl:sim1mae} presents summary statistics pertaining to the recovery of the basis functions weights averaged across levels, partitions, and the 50 simulations. As the table indicates, the true values of the basis functions weights are contained in the 95\% credible intervals with a frequency that is close to the nominal level, at times slightly over. 
The accuracy with which the basis functions weights are estimated varies depending on the magnitude of the basis functions weights. While on average across simulations, the average relative absolute error is around 0.40, when $L$ = 100, the average relative absolute error of basis functions weights whose true absolute value is less than 0.5 is 1.54, while for basis functions weights with true absolute value greater or equal than 0.5, it is significantly smaller and equal to 0.28.
We also observe that, when $L$=10,000, the MCMC algorithm never samples basis functions weights from  the shrinkage prior resulting in a larger average Mean Absolute Error (MAE) and Mean Square Error (MSE). On the other hand, $L$=100 guarantees that the MCMC algorithm samples the zero-valued basis functions weights from the shrinkage prior, leading to a greater accuracy in recovering the $\boldsymbol{\eta}_{m,j}$'s. 

\subsubsection{Simulation Study 2}\label{sec:sim23res} 
Data were generated at $n=1,012$ random locations in $\mathcal{S}=[0,1]\times [0,1]$ according to the following model:
 \begin{equation}
 y(\mathbf{s}) = \mu(\mathbf{s}) + I(\mathbf{s}_{x} < 0.5) w_{1}(\mathbf{s}) + I(\mathbf{s}_{x} \geq 0.5) w_{2}(\mathbf{s}) +  \epsilon(\mathbf{s}) \qquad \epsilon(\mathbf{s}) \stackrel{iid}{\sim} N(0,\tau^2)
 \label{eq:simdata}
 \end{equation}
 where $\mathbf{s}_{x}$ indicates the first coordinate of the two-dimensional vector of geographical coordinates for point $\mathbf{s}$ (e.g. longitude or Easting). In (\ref{eq:simdata}),  $\tau^2$ was set equal to $0.05$, and $w_{1}(\mathbf{s})$ and $w_{2}(\mathbf{s})$ were taken to be two mean-zero stationary Gaussian processes with  
 Mat\'{e}rn covariance functions with parameters $\sigma^2=1$, $\nu=1$, and $\phi$ equal to $1.0$ and $ 0.01$, respectively. 
 
We selected 756 locations at random and we used the corresponding data as training data. To these data, we fitted: (i) the mixture M-RA model with a stationary Mat\'{e}rn covariance function to define the basis functions, and (ii) the M-RA model of \citet{Katzfuss2017} without shrinkage of the basis functions weights. The comparison with the M-RA model will enable us to determine whether shrinkage is needed to determine the regions of inhomogeneities (goal (2)). In fitting the mixture M-RA model, we used $M=3$, $J=4$, and $r=16$, with $L$ determined via tuning and kept equal to its optimal value ($L$=100) in the post burn-in iterations. 

\begin{figure}
\begin{center}
\subfigure[]{ \includegraphics[scale=0.42]{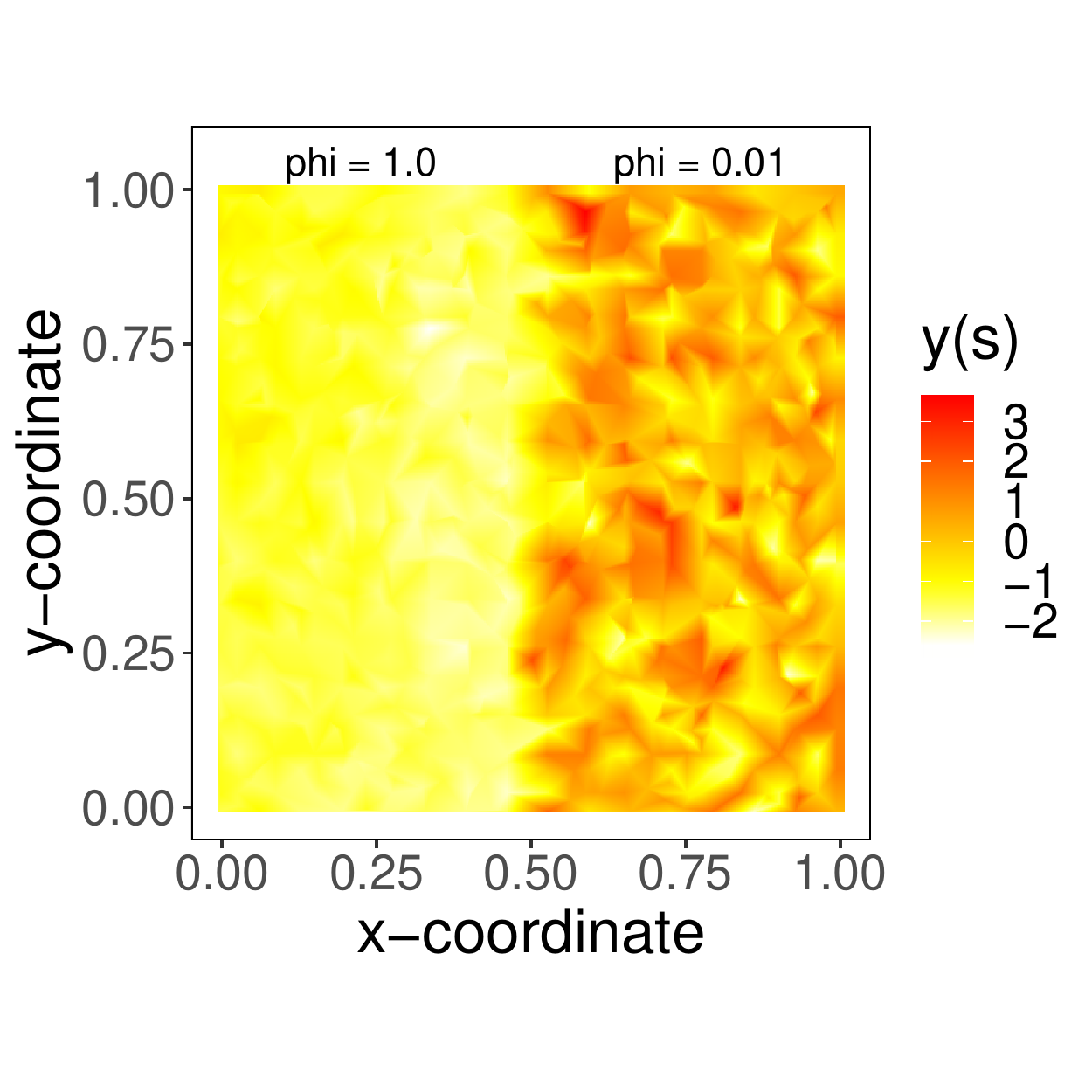}} \\ 
\subfigure[]{\includegraphics[scale=0.2]{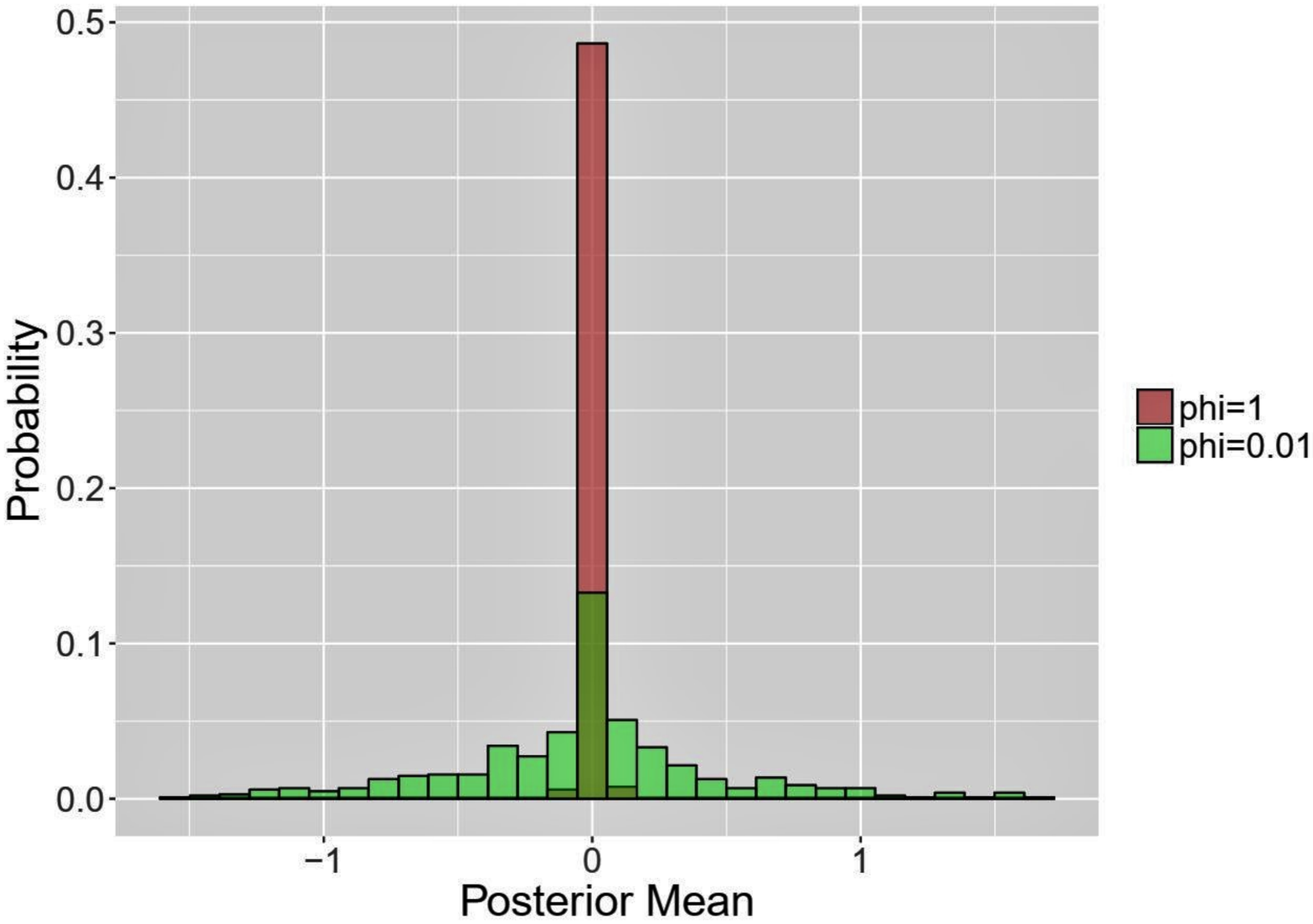}} \qquad
\subfigure[]{\includegraphics[ trim=0 100 0 0, clip, scale=0.2]{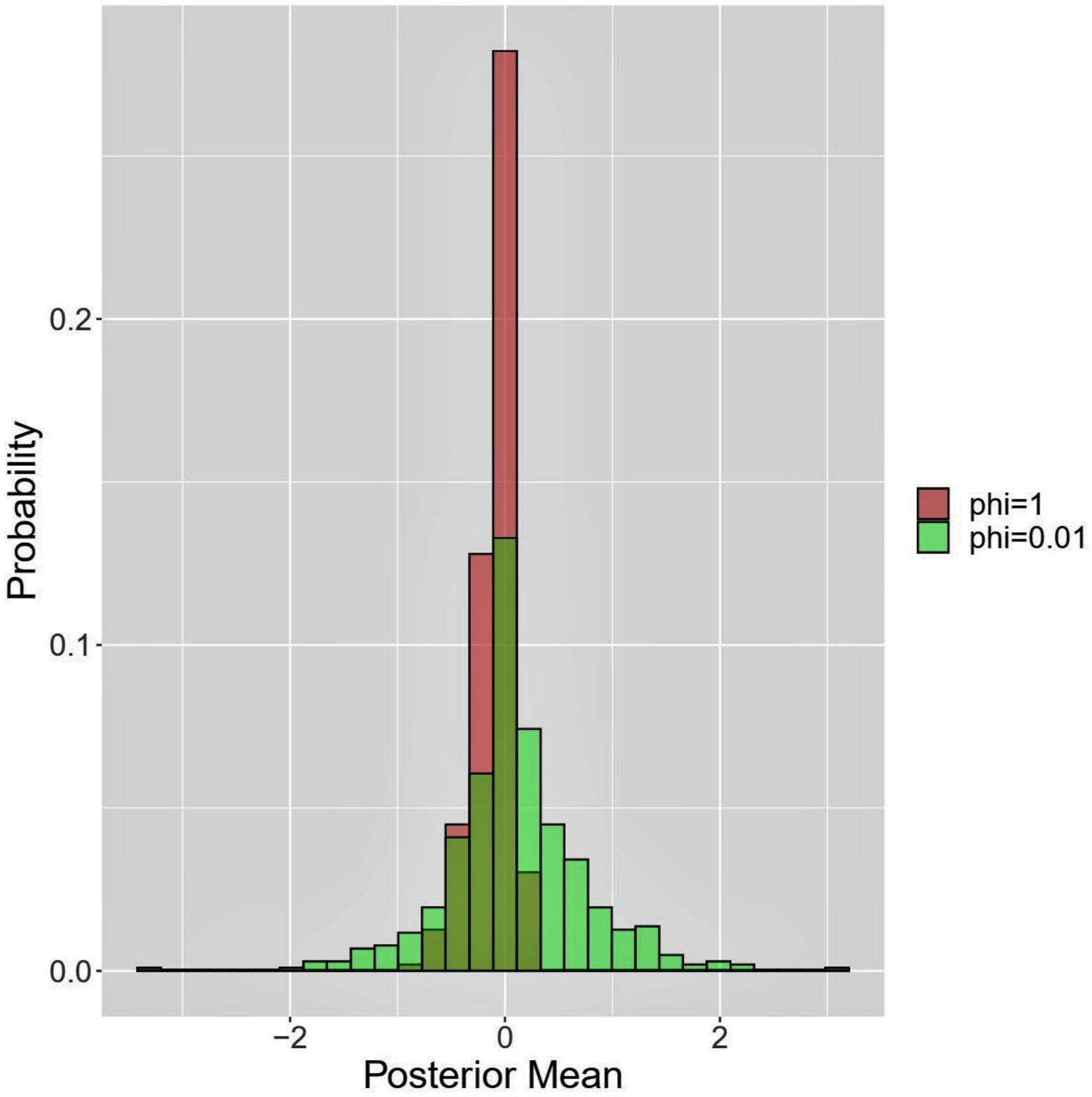}} \\
\subfigure[]{\includegraphics[scale=0.42]{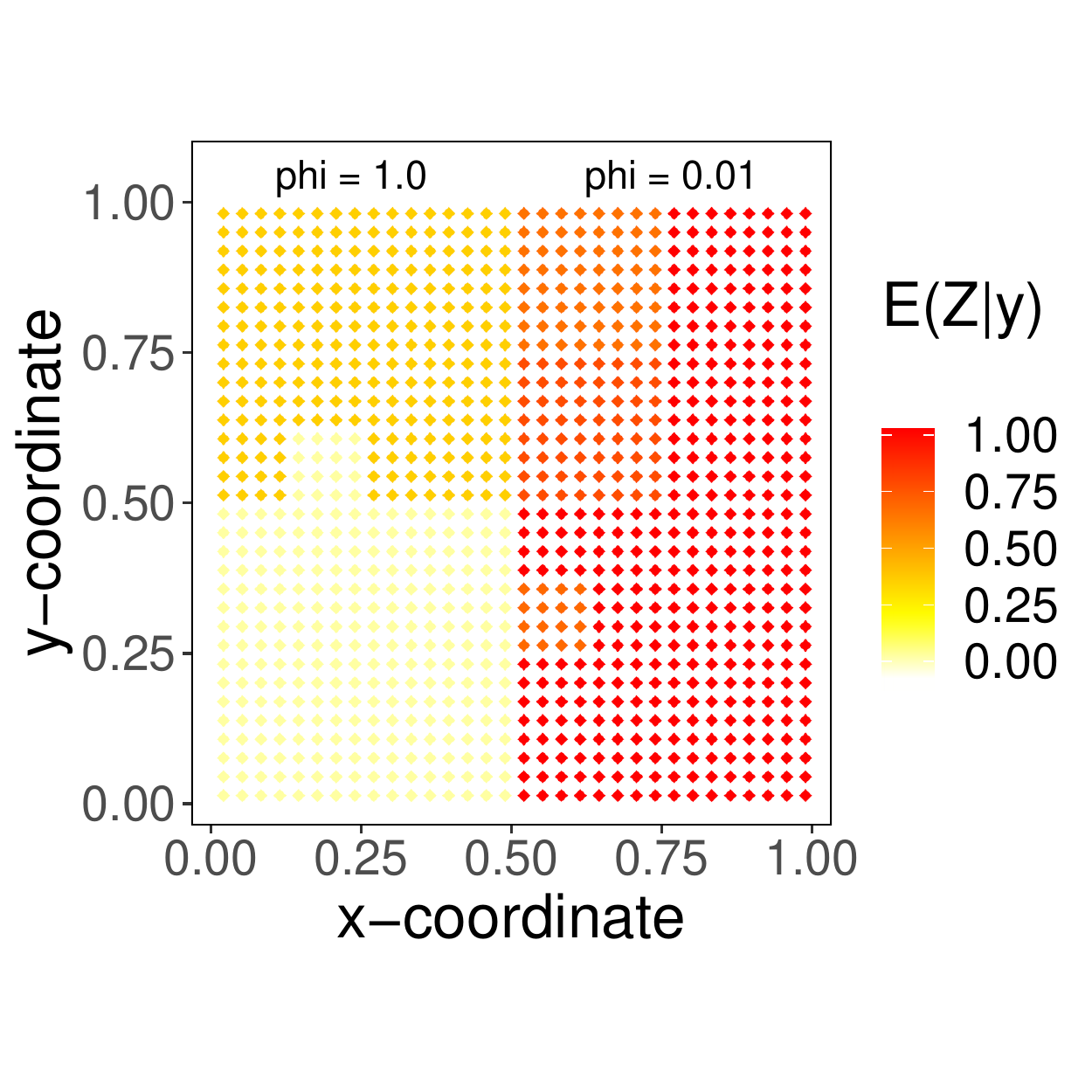}} 
\subfigure[]{\includegraphics[scale=0.42]{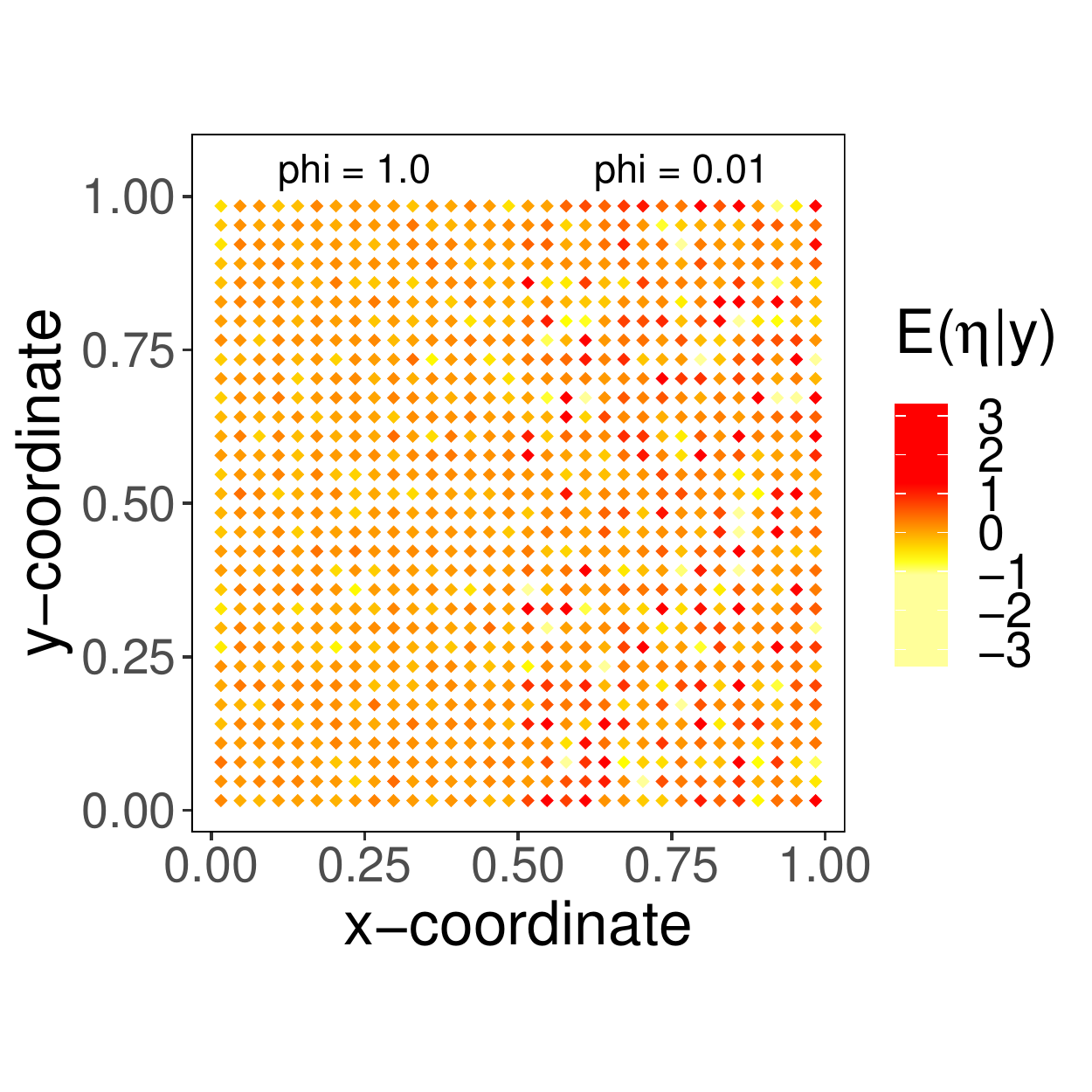}}
\end{center}
\caption{Simulation study 2. (a) One of the 30 realizations of $y(\mathbf{s})$ generated according to (\ref{eq:simdata}), with $\mu(\mathbf{s}) \equiv 0$, $\forall \mathbf{s} \in \mathcal{S}=[0,1] \times [0,1]$, $\tau^2=0.05$, and $w_{1}(\mathbf{s})$ and $w_{2}(\mathbf{s})$ mean-zero stationary Gaussian processes with Mat\'{e}rn covariance function with parameters, $\sigma^2_1=1.0$, $\nu_1=1$, $\phi_1=0.1$ and  $\sigma^2_2=1.0$, $\nu_2=1$, and $\phi_2=1.0$, respectively. (b)-(c) Histograms of the posterior means of the basis function weights $\boldsymbol{\eta}_{m,j}$ for $m=3$ for the mixture M-RA and the M-RA model without shrinkage, respectively, grouped by values of $\phi$, the range parameter. (d) Posterior means of the $Z_{m,j}$'s for $m=3$. (e) Posterior means of the basis function weights $\boldsymbol{\eta}_{m,j}$ for $m=3$ as estimated by the M-RA model without shrinkage.
 \label{nsillus}}
 \end{figure}

Inspecting the results for the basis functions weights for simulation study 2, we observe that the $\boldsymbol{\eta}_{m,j}$'s are sampled from the two components of the mixture prior at different rates in the two halves of the spatial domain: in the part of the domain where $\phi = 1.0$, the average posterior mean of the binary latent variables $Z_{m,j}$ at the third level ($m=3$), averaged across the 30 simulations, is 0.296. Conversely, in the region where $\phi = 0.01$, it is 0.968. Also the M-RA model without shrinkage captures the difference in magnitude between the basis functions weights in the two regions. However, looking at the spatial map of the posterior means of the basis functions weights displayed in Figure~\ref{nsillus}(e), it is clear that the M-RA cannot identify the two regions easily.  

A similar conclusion is drawn when examining the confusion matrix presented in Table~\ref{table:confuse}. This table is obtained by classifying the knots in the highest level ($m$=3) of the mixture M-RA and of the M-RA to a region according to some model-specific criteria. For the mixture M-RA, since the binary latent variables are defined at the partition level, all knots in partition $j$ at level $m=3$ are classified as belonging to Region 1 ($\phi=1.0$) or Region 2 ($\phi=0.01$), based on whether $E\left[ Z_{m,j} | \mathbf{y} \right] < 0.5$ or not. This is also referred to as the Median Probability Model of \citet{barbieri&2004}. On the other hand, in the M-RA model, a knot at level $m=3$ is labeled as belonging to Region 1 or Region 2 based on whether the absolute value of the posterior mean of the basis function weight associated with the given knot is below a certain threshold. The results presented in Figure~\ref{nsillus} and Table~\ref{table:confuse} underscore the importance of the shrinkage prior in detecting regions of non-stationarity (goal (3)).


\subsection{Analysis of Soil Organic Carbon in Continental US}\label{sec:SOC}
\begin{figure}
    \centering
    \subfigure[]{\includegraphics[width=0.45\textwidth]{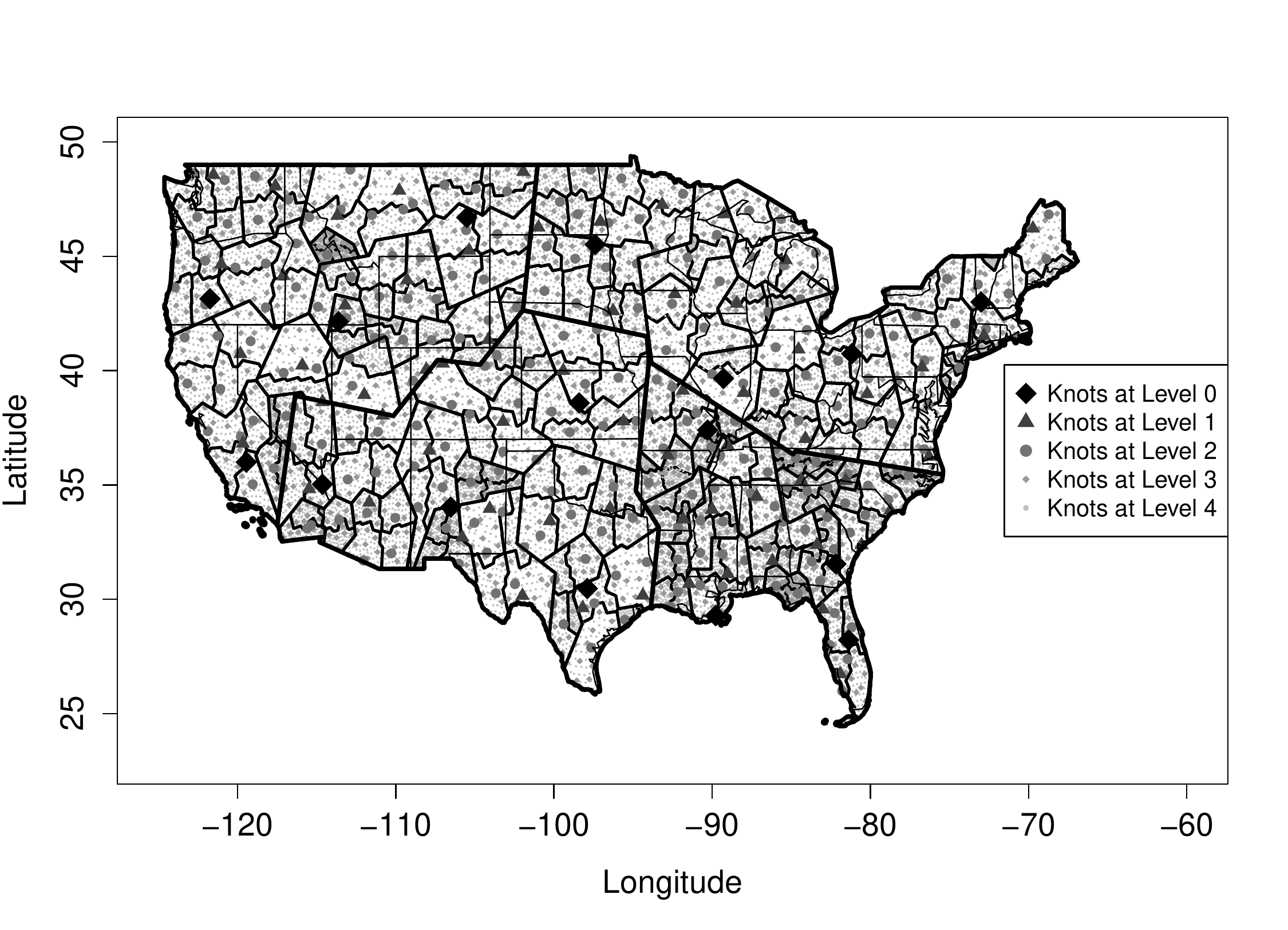}}  \qquad
   \subfigure[]{\includegraphics[width=0.45\textwidth]{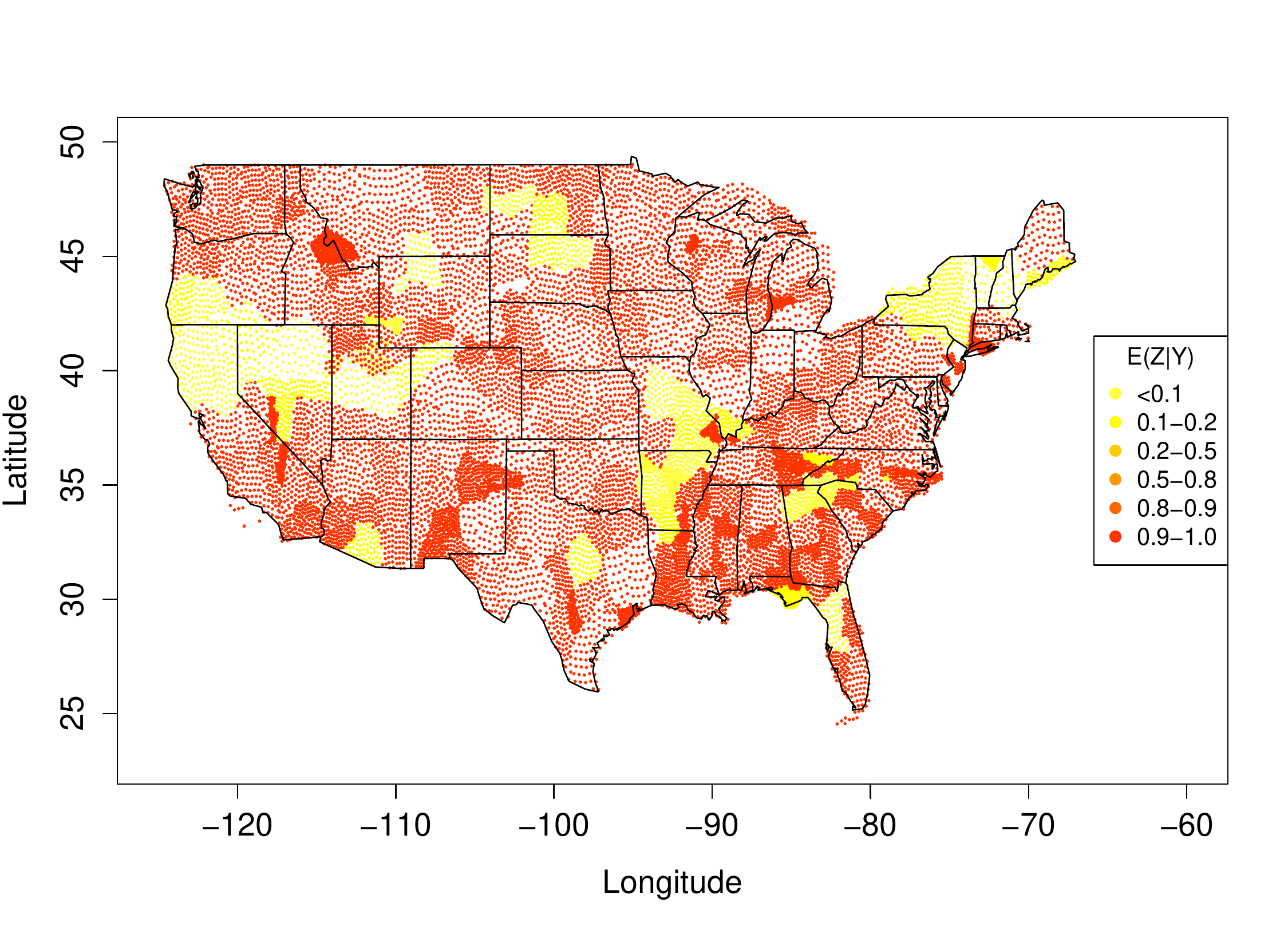}}   \\
    \subfigure[]{\includegraphics[width=0.45\textwidth]{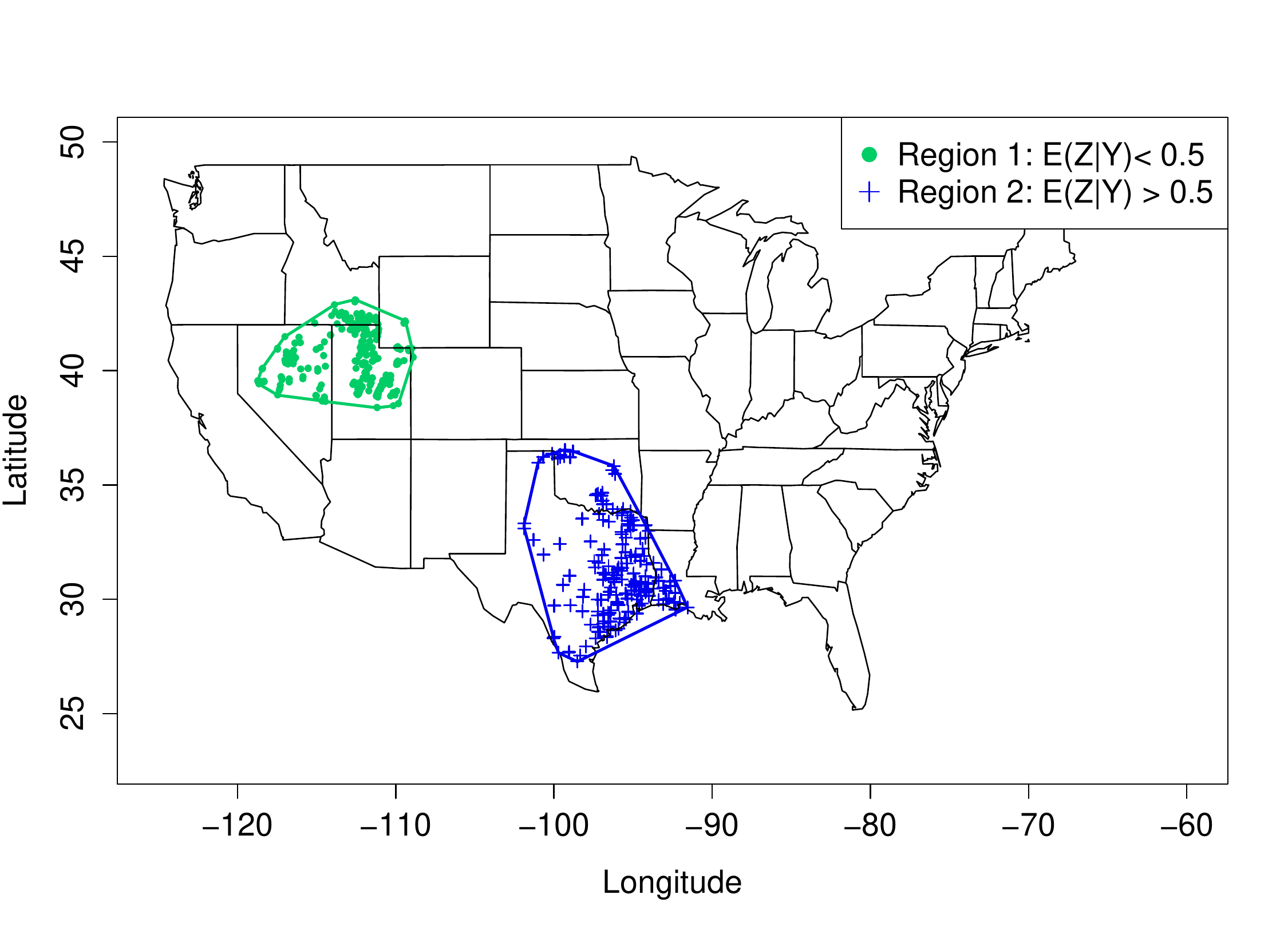}} \qquad
    \subfigure[]{\includegraphics[width=0.35\textwidth]{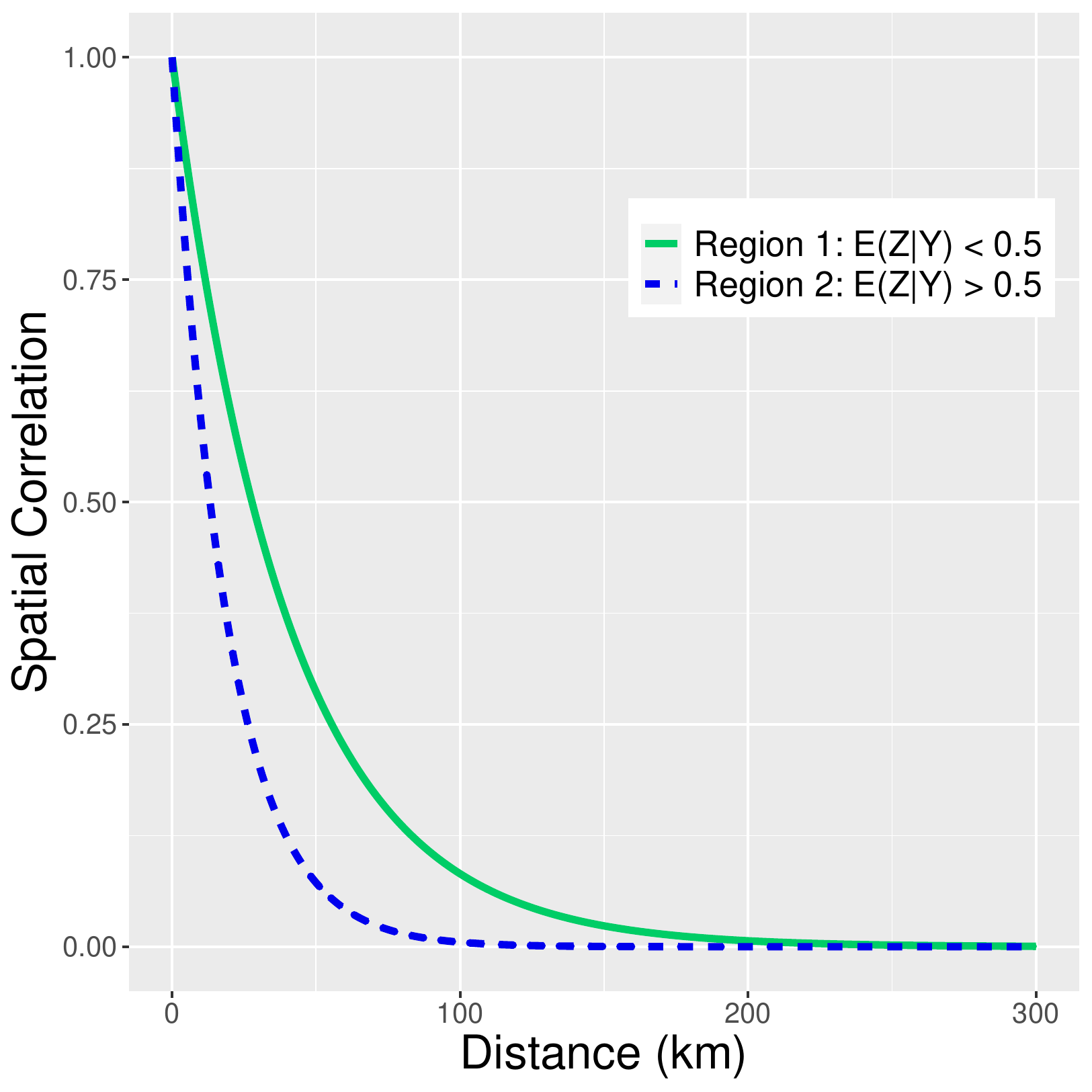}} \\
        \subfigure[]{\includegraphics[width=0.4\textwidth]{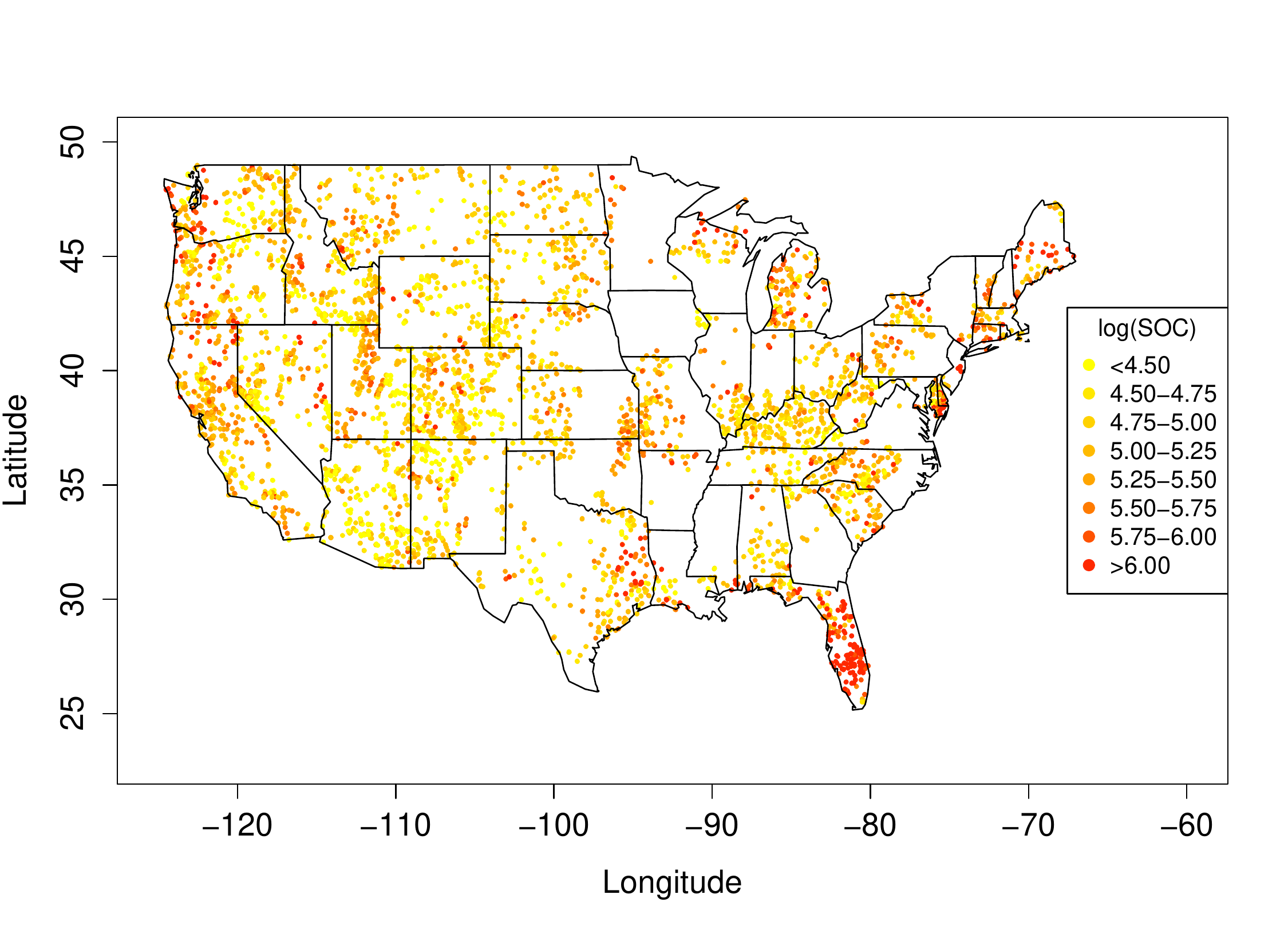}} \qquad
    \subfigure[]{\includegraphics[width=0.4\textwidth]{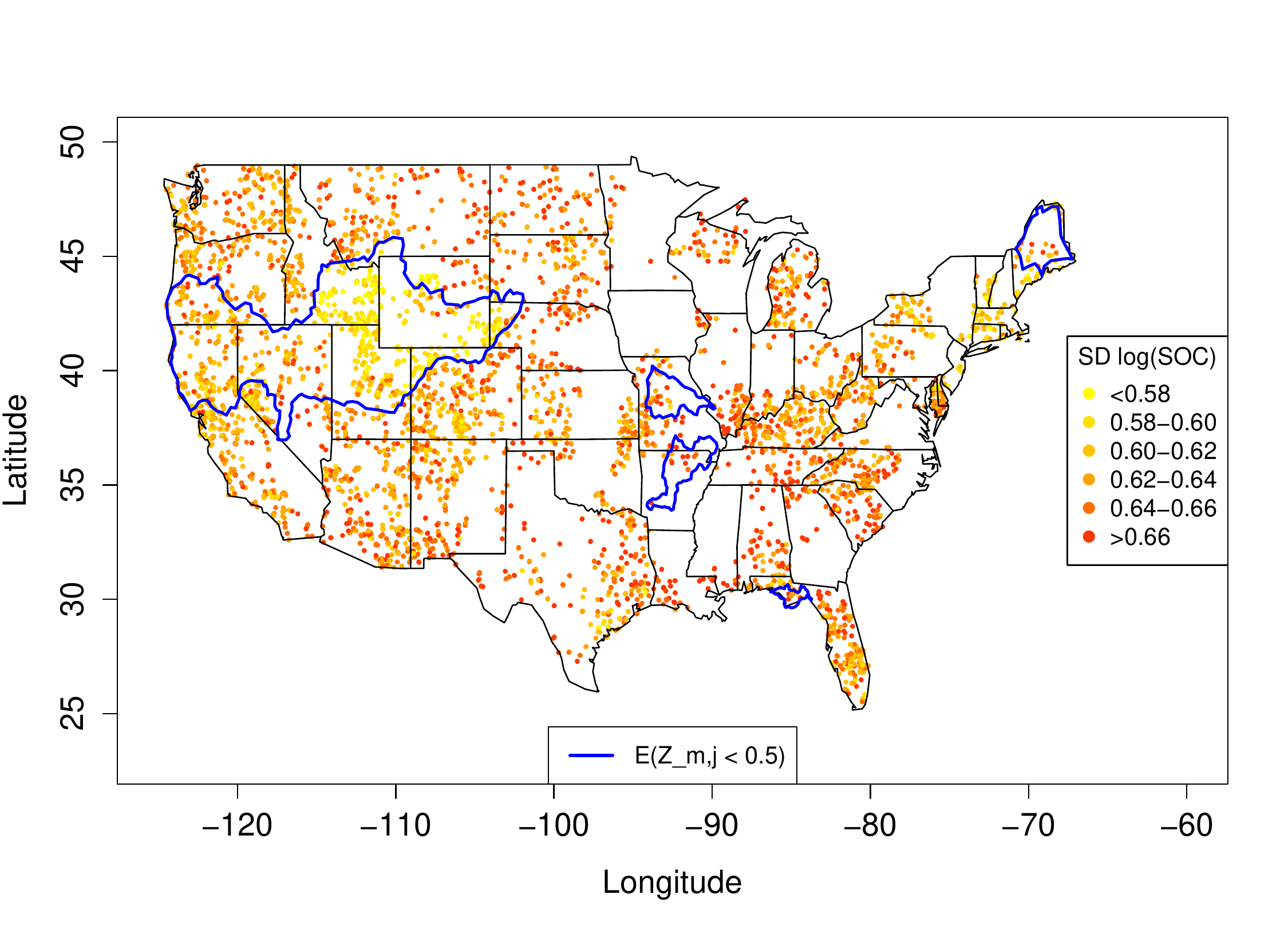}} 
    \caption{SOC analysis. (a) Placement of the knots used in the mixture M-RA model. (b) Posterior means of $Z_{m,j}$ at the highest level ($m$=4). (c) Two regions where the posterior means of $Z_{m,j}$ at the highest level ($m$=4) are, respectively, less than 0.5 versus greater or equal to 0.5. In the former, basis functions weights are more likely to be shrunk to zero while in the latter they are more likely to not be  shrunk to zero.  (d) Estimated Mat\'{e}rn correlation functions in the selected subregions. (e) Predicted SOC and (f) posterior predictive standard deviation as yielded by the mixture M-RA model. In (e), lines delineate regions where the posterior mean of the $Z_{m,j}$'s at the highest level, $m$=4, is less than 0.5. We identify these as regions of local stationarity.}
    \label{figure:postzSOC}
\end{figure}

To the 20,087 measurements of $\log$ SOC described in Section~\ref{sec:data} we fit our mixture M-RA model in (\ref{eq:priorweights2}), with $\mu(\mathbf{s})$ a linear function of land use/land cover, drainage class and elevation. We used a stationary Mat\'{e}rn covariance function to define the basis functions, and set $M$, $J$, and $r$ equal to 4, 4 and 16, respectively. To allow for regions of local stationarity with irregular boundaries, we placed knots using a Voronoi tessellation approach: more discussion on this approach is provided in Web Appendix E, where we also lists all the prior specifications (see Web Table 8). We ran the MCMC algorithm for 10,000 iterations, determining the optimal value of $L$ in the burn-in period ($L=100$) and keeping it fixed afterwards. We assessed convergence using multiple diagnostics, which we present in Web Appendix E.
After discarding the first 5,000 iterations for burn-in, we derived posterior inference on the latent variables $Z_{m,j}$ which we use to identify regions where the basis functions weights are shrunk to zero versus not shrunk. 
Figure~\ref{figure:postzSOC}(b) plots the posterior means of the latent binary variables $Z_{m,j}$ at their respective knots locations.  The regions in which the posterior means of the $Z_{m,j}$'s are below 0.5 are the regions where the basis functions weights are shrunk to zero, and those are expected to contain observations whose residual spatial correlation decays more slowly. To validate this, using land use/land cover, drainage class, and elevation as covariates, we fit a likelihood-based spatial statistical model to SOC in the two regions denoted in Figure \ref{figure:postzSOC}(c): Region 1, further West, which contains knots corresponding to basis functions weights shrunk towards zero, and Region 2, SouthEast of Region 1, in which basis functions weights are not shrunk.  Figure \ref{figure:postzSOC}(d) shows the estimated Mat\'{e}rn correlation function in the two regions.  Consistent with our expectation, Region 1, in which basis functions weights are shrunk towards zero, exhibits a more slowly-decaying spatial correlation than Region 2, in which the basis functions weights are active in the model.

Using our mixture M-RA model, we also generate predictions of SOC at observation sites along with the corresponding posterior predictive standard deviations, which we display in Figure~\ref{figure:postzSOC}(e) and \ref{figure:postzSOC}(f). Combining the information in panels (e) and (f) of Figure~\ref{figure:postzSOC} together allows us to identify regions where the residual spatial correlation in SOC persists at long distances and predictions are characterized by great uncertainty. This can in turn be used to plan future SOC data collection campaigns. Specifically: more sampling efforts should be concentrated in regions with large prediction uncertainty and where the spatial correlation has a short effective range. Based on Figure~\ref{figure:postzSOC}, this means: Texas, Montana, Western North and South Dakota, and Northern Mississippi/Alabama/Georgia, among others.

As a secondary assessment of our model, we also examined its predictive performance at 2,000 hold-out sites. The goal of this added assessment was two-fold: to obtain an indication of the adequacy of our model for spatial interpolation, and to evaluate its appropriateness for modeling SOC. For this latter reason, we compare its predictive accuracy against that of three other models: (i) a stationary Bayesian Kriging model that we fit using a predictive process approximation \citep{Banerjee&2008} due to the large size of the dataset; (ii) the M-RA model of \citet{Katzfuss2017}; and (iii) the convolution-based non-stationary model of \citet{RisserCalder}. Among all the non-stationary models proposed in the literature, we select the latter partly because statistical software to implement it is readily available. Results relative to this out-of-sample predictive performance assessment are reported in Web Appendix E. Here, we simply observe that although our model does not consistently yield the best predictions in terms of Mean Square Predictive Error (MSPE) or Mean Absolute Predictive Error (MAPE), it has similar predictive accuracy than the original M-RA and the model of \citet{RisserCalder}, and it yields much better predictions than the stationary Bayesian Kriging model which has the worst prediction accuracy of all. These results indicate that even though our model has been developed with the main purpose of identifying regions of non-stationarity, it also has a utility beyond that: it is a viable non-stationary spatial statistical model in its own right.  

\section{Discussion}\label{sec:discussion}
Analysis of geostatistical data often involves as a first step the selection of a covariance function. In applications, stationary covariance functions are often used even though the assumption of stationarity might not be warranted. In this paper, we have proposed a flexible modeling framework that allows investigators to gain a better understanding of the spatial dependence structure of a spatial process while accomodating both stationary and non-stationary spatial data, in cases where the non-stationarity is due to inhomogeneities in the range of the spatial correlation. Application of our model to both stationary (see simulation study 5 in Web Appendix D) and non-stationary data have shown that our model allows one to identify regions of local stationarity, if they exist. 

Determining whether there exist regions in the spatial domain where a spatial process displays varying strength of spatial dependence is extremely important for various reasons. 
From a sampling design perspective, knowing that in different regions the spatial process is characterized by a different range parameter, could lead to a differential strategy when collecting observations or when placing monitoring devices, as we have discussed in the analysis of SOC in Section~\ref{sec:SOC}.
From a computational point of view, the decomposition of the spatial domain in regions of local stationarity can lead to computational savings as a spatial model can be fit to data within each region individually.

Another appealing feature of our model is that it can identify regions of local stationarity by simply using a stationary covariance function within the M-RA without the need of using a non-stationary covariance function model, for which parametric closed form expressions are not readily available. 

There are multiple ways in which our model could be extended and improved. First, for now we have only been considering Gaussian spatial processes: it would be interesting, and potentially not too difficult, to extend the mixture M-RA modeling framework to non-Gaussian spatial data.  
In its current form, our model only accommodates non-stationarity due to inhomogeneities in the range of the correlation function; extensions of this work could be geared towards accommodating other types of non-stationarity. The use of an anisotropic spatial covariance function could also be explored.

The prior specification on the M-RA basis functions weights involves a mixture of two normal priors, with one of the two normal distributions introducing an additional parameter, the shrinkage parameter $L$. In our implementation, $L$ is tuned during the burn-in period and kept it fixed afterwards. In all our analyses the optimal value of $L$ resulted to be 100: we suspect that the magnitude of $L$ is influenced by the size of the marginal variance, as in all our simulations and in the SOC data, the marginal variance was approximately equal to 1.0. A further avenue of research could be to investigate how to provide a prior on $L$, and infer upon it using the data. 

In specifying the mixture prior on the basis functions weights, we used shrinkage probabilities that vary by level, and have expressed them as powers of a parameter $\rho$. A more general formulation would allow the shrinkage probabilities to vary from level to level and not be multiple of a common parameter $\rho$. Although this approach might increase the degrees of freedom of the model, it might not ensure that the shrinkage probabilities decrease with increasing level, as one would expect. 

In our simulation experiments, we have placed the knots on a regular grid or at the observation locations in the highest level of the M-RA (see simulation studies in Web Appendix D), yielding regions of local stationarity that have regular boundaries. We note that it is possible to identify regions with more irregular boundaries by placing the knots at random, defining the M-RA subregions as disjoint unions of complete Voronoi polygons, as we show in Web Appendix A. We have used this strategy in the analysis of SOC. Investigating the influence of the knots' placement on the inference yielded by the mixture M-RA is another avenue for future research.

Finally, we observe that while prediction is not the main goal of our model, the predictive performance of the mixture M-RA in both simulations (see Web Appendix (D)) and in the SOC data analysis, is either comparable or slightly inferior to that of the MRA model or another non-stationary model. While this is a limitation of our model to be further improved, we underline that the primary goal of our model is to capture the second-order structure of a spatial process.


\bibliographystyle{apa} 
\bibliography{mixture_MRA_Jan2021}  

\section*{Supporting Information}
Web Appendices, Tables, and Figures referenced in Sections 3.1, 3.3, 4.1, 4.2 and 5, are available in a supplemental document submitted as an ancillary file. Code for implementing the Mixture M-RA to simulated data are available at:\\ \texttt{https://github.com/marco-benedetti/Mixture-M-RA}.

\section*{Data Availability Statement}
The data that support the findings of this study are openly available in the \texttt{soilDB} package of \texttt{R} at \texttt{https://cran.r-project.org/web/packages/soilDB/index.html}. 

\label{lastpage}


\clearpage

 \clearpage

\clearpage

\clearpage

\begin{table}[!h] 
\centering 
  \caption{Exploratory data analysis for SOC. Estimated regression coefficients with corresponding standard errors from a linear regression model that regresses log(SOC) on land use/land cover, drainage class, and elevation. $^{***}$ denotes estimates significant at $\alpha=0.001$, while $^{**}$ indicates significance at $\alpha=0.05$  level. \label{tbl:linmodSOC}} 
\begin{tabular}{rlcc} 
\hline  
Variable & Level & $\hat{\beta}$ & $SE\left( \hat{\beta} \right)$ \\ 
\hline \\[-1.8ex] 
Intercept &  & 5.017$^{***}$   & 0.021 \\ 
Land use/land cover & Farmland & 0.198$^{***}$  & 0.021 \\ 
\small{(Reference = Crop)} & Pastureland & 0.141$^{***}$  & 0.023 \\ 
& Range & $-$0.205$^{***}$  & 0.021 \\ 
& Wetland & 0.081$^{**}$  & 0.029 \\ 
& CRP & 0.049     & 0.035 \\ 
Drainage Class&  Very poorly drained & 1.695$^{***}$  & 0.035 \\ 
\small{(Reference = Not rated)} & Poorly drained & 0.766$^{***}$   & 0.028 \\ 
& Somewhat poorly drained & 0.335$^{***}$ & 0.027 \\ 
& Moderately well-drained & 0.169$^{***}$ & 0.026 \\ 
&  Well-drained & 0.048$^{**}$ & 0.017 \\ 
& Somewhat excessively drained & 0.182$^{***}$   & 0.039 \\ 
& Excessively drained & 0.343$^{***}$ & 0.046 \\ 
Elevation & & $-$0.0002$^{***}$  & 0.00001 \\ 
\hline 
\multicolumn{4}{l}{Estimated residual 
standard error: $0.88 \log \left( \mbox{Mg C ha}^{-1} \right)$}\\
\hline 
\end{tabular} 
\end{table} 

\clearpage

\begin{table}[!h]
\begin{center}
\caption{Simulation study 1. Average posterior means of the $Z_{m,j}$, average Mean Absolute Error (Avg. MAE), average Mean Squared Error (Avg. MSE), average bias, average relative MSE, and average empirical coverage (covg.) of the 95\% credible interval (CI) for the non-zero basis functions weights, averaged across levels, subregions, and the 50 simulated datasets. Summary statistics are presented overall, and stratified based on whether the true basis functions weights are equal to zero or not. 
\label{tbl:sim1mae}}
\begin{tabular}{|c|c|c|c|c|c|c|c|c|c|}
\hline
                                           &                &                   &                          &                                &                      &     Avg.            & Avg. & Avg.        & Avg.   \\
                                           &                &   Avg.  & Avg.             & Avg.                  & Avg.       &   Relative & Relative & Relative        &  covg. of
  \\
$\boldsymbol{\eta}_{m,j}$  & $L$         &   $E \left[ Z_{m,j}| \mathbf{y} \right]$  & MAE                  & MSE                        & bias                & MAE & bias &  MSE      &  95\% CI \\
\hline
                                                  &    10       &  0.476 &  1.214  &   5.379  & - 0.021   & NA  & NA & NA & 0.912 \\  
                                                  &    25       &  0.518  & 1.227  &   5.394  & - 0.019   & NA  &  NA & NA & 0.925 \\  
                                                 &    50       &  0.536 &  1.244  &   5.415  & - 0.017   &  NA&  NA & NA &0.926 \\ 
  All                                                &  100      &0.600 &  1.341   &  5.453   & -0.023    &  NA &  NA & NA &0.935 \\
                                                 & 200      &   0.828 & 1.607   &  5.952   &  -0.023   &  NA &  NA & NA &0.930 \\
                                                 & 1,000      &   0.884 & 1.661  &  6.029   &  -0.022  &  NA &  NA & NA &0.930 \\
                                                  & 10,000  &  1.000 &  1.687   &  6.131   & -0.024    & NA &  NA & NA & 0.921 \\
\hline
                                                   &   10        &   0.000 &  0.041    & 0.003    &   -0.000   & NA &  NA & NA & -- \\ 
                                                   &   25        &  0.001 &   0.041    & 0.003    &   -0.000   & NA &  NA & NA & -- \\ 
                                                   &   50        & 0.025 &    0.062    & 0.009    &   -0.000   & NA &  NA & NA & -- \\ 
 $ = 0$   &  100      &   0.152 &  0.302    & 0.377     &  -0.005    & NA &  NA & NA & -- \\
                                                   &  200      &   0.634&  0.874    & 1.535    &  -0.008    &  NA &  NA & NA & -- \\
                                                   &  1,000      &   0.754&  0.912    & 1.719    &  -0.009    &  NA &  NA & NA & -- \\
                                                   &  10,000  &  1.000 &  1.043    & 1.909    &  -0.010    & NA &  NA & NA & -- \\
\hline
                                                   &   10        &  0.904 &   2.356    & 10.914   &   -0.033   & 0.491 & $-$0.010 & 0.519 &   0.831 \\ 
                                                   &   25        &  0.978 &   2.301    & 10.493   &   -0.032   & 0.476 & $-$0.008  & 0.491 & 0.866 \\ 
                                                   &    50         &  0.991 &  2.295      &  10.221  &  -0.032    & 0.401 & $-$0.008  & 0.494& 0.861 \\ 
$ \ne 0$ & 100       & 0.999 &   2.264      &  9.966    &  -0.039    & 0.440 & $-$0.011  & 0.481 &0.927 \\
                                                   &  200       &  1.000 &  2.258      &  9.879    &  -0.037    & 0.488 & $-$0.011  & 0.507 &0.949 \\
                                                   &  1,000       &  1.000 &  2.259      &  9.887    &  -0.036    &0.489 & $-$0.011  & 0.507 &0.950 \\
                                                   &  10,000   & 1.000 & 2.260      &  9.884    &  -0.037    & 0.450 & $-$0.011  & 0.503 &0.949\\
\hline
\end{tabular}
\end{center}
\end{table}

\clearpage

\begin{table}[!p]
    \centering
     \caption{Confusion matrix for simulation study 2 relative the M-RA model without shrinkage and the mixture M-RA. In the M-RA model without shrinkage, classification of knots $h$ in partition $j$ of level $m$ ($m$=3 here) is determined based on whether $| E \left( \mathbf{\eta}_{m,j,h} | \mathbf{y} \right) | < t$ or not, with $t$ denoting a threshold. If $| E \left( \mathbf{\eta}_{m,j,h} | \mathbf{y} \right) | < t$, then the $h$-th knots is classified as belonging to Region 1, otherwise to Region 2. The classifier for the mixture M-RA model is based on the median probability model: all knots in parition $j$ of level $m$ ($m=3$) are classified to belong to region 1 if $E \left( Z_{m,j} | \mathbf{y} \right) < 0.5$, whereas they are classified as belonging to Region 2 otherwise.
      \label{table:confuse}}
   \begin{tabular}{|c||c||c|c|c|}
    \hline
  &  Threshold   & &  Region 1  & Region 2     \\
  &   $t$  & &  ($\phi = 1.00$) & ($\phi=0.01$)  \\ \hline
  \multirow{20}{*}{M-RA}  &  \multirow{4}{*}{0.1} &   Classified as Region 1 & 4,919 & 2,452   \\ 
   &     & Classified as Region 2 & 10,441 & 12,908   \\   
    &   &   Percent Correctly & &   \\
    &   &  Classified & \textbf{32.0\%} & \textbf{16.0\%} \\  
        \cline{2-5} 
    & \multirow{4}{*}{0.25} &  Classified as Region 1 & 10,328 & 5,758   \\ 
     &   &    Classified as Region 2 & 5,032 & 9,602 \\ 
     &   &  Percent Correctly & &  \\
    &   &  Classified & \textbf{67.2\%} & \textbf{62.5\%}  \\  
         \cline{2-5}
    &  \multirow{4}{*}{0.5} & Classified as Region 1 & 13,916 & 9,604 \\  
    &  &  Classified as Region 2 & 1,444 & 5,756 \\  
    &   & Percent Correctly & & \\
     &  & Classified & \textbf{90.6\%} & \textbf{37.4\%} \\ 
        \cline{2-5}
    & \multirow{4}{*}{1.0}  & Classified as Region 1 & 15,100 & 13,291 \\   
    &  &  Classified as Region 2 & 260 & 2,069 \\   
    &  &  Percent Correctly & & \\
    & &  Classified & \textbf{98.3\%} & \textbf{13.5\%}  \\  
        \cline{2-5}
    &  \multirow{4}{*}{1.5}    & Classified as Region 1 & 15,316 & 14,677 \\ 
    &   & Classified as Region 2 & 44 &  683  \\  
    &   & Percent Correctly & & \\
    &  &   Classified & \textbf{99.7\%} & \textbf{4.4\%} \\ 
        \hline  \hline
        \hline  \hline
 \multirow{6}{*}{Mixture M-RA}    &   &  $E(Z_{m,j} | \mathbf{y})  < 0.5$  &  &   \\
    &  &  Classified as Region 1 & 12,800 & 0 \\ 
    & &    $E(Z_{m,j} | \mathbf{y})  \ge 0.5$ & & \\
  &    & Classified as Region 2 & 2,560 &  15,360  \\  
    &  &   Percent Correctly & & \\
    &   & Classified &  \textbf{83.3\%} & \textbf{100.0\%}  \\
     \hline 
    \end{tabular}
\end{table}

\clearpage

\end{document}